\documentclass{acm_proc_article-sp}

\begin{document}

\title{pSpatiocyte: A Parallel Stochastic Method for Particle Reaction-Diffusion Systems}

\numberofauthors{4}
\author{
\alignauthor Atsushi Miyauchi\titlenote{Current address: miyauchi@rist.or.jp.}\\
\affaddr{HPCI Program for Computational Life Sciences\\ RIKEN}\\
\affaddr{7-1-26 Minatojima-minami}\\
\affaddr{Chuo, Kobe, Japan}\\
\email{atsushi.miyauchi@riken.jp}
\alignauthor Kazunari Iwamoto\\
\affaddr{Center for Integrative Medical Sciences\\ RIKEN}\\
\affaddr{1-7-22 Suehiro-cho}\\
\affaddr{Tsurumi, Yokohama, Japan}\\
\email{kiwamoto@riken.jp}
\alignauthor Satya Nanda Vel Arjunan\\
\affaddr{Quantitative Biology Center \\ RIKEN}\\
\affaddr{6-2-3 Furuedai}\\
\affaddr{Suita, Osaka, Japan}\\
\email{satya@riken.jp}\\
\and 
\alignauthor Koichi Takahashi\\
     \affaddr{Quantitative Biology Center\\ RIKEN}\\
     \affaddr{6-2-3 Furuedai }\\
      \affaddr{Suita, Osaka, Japan }\\
      \email{ ktakahashi@riken.jp}
}

\maketitle


\begin{abstract}
Computational systems biology has provided plenty of insights into cell biology. Early on, the focus was on reaction networks between molecular species. Spatial distribution only began to be considered mostly within the last decade. However, calculations were restricted to small systems because of tremendously high computational workloads. To date, application to the cell of typical size with molecular resolution is still far from realization. In this article, we present a new parallel stochastic method for particle reaction-diffusion systems. The program called pSpatiocyte was created bearing in mind reaction networks in biological cells operating in crowded intracellular environments as the primary simulation target. pSpatiocyte employs unique discretization and parallelization algorithms based on a hexagonal close-packed lattice for efficient execution particularly on large distributed memory parallel computers. For two-level parallelization, we introduced isolated subdomain and tri-stage lockstep communication for process-level, and voxel-locking techniques for thread-level. We performed a series of parallel runs on RIKEN's K computer. For a fine lattice that had relatively low occupancy, pSpatiocyte achieved 7686 times speedup with 663552 cores relative to 64 cores from the viewpoint of strong scaling and exhibited 74\% parallel efficiency. As for weak scaling, efficiencies at least 60\% were observed up to 663552 cores. In addition to computational performance, diffusion and reaction rates were validated by theory and another well-validated program and had good agreement. Lastly, as a preliminary example of real-world applications, we present a calculation of the MAPK model, a typical reaction network motif in cell signaling pathways.
\end{abstract}





\section{Introduction}
Computational cell biology is a relatively new and rapidly growing field that uses modern computers and computational science to study microscopic biological phenomena. Its primary objective is to quantitatively understand cellular behavior at the molecular level. Because a large number of physical and chemical processes simultaneously take place in a cell, approximation methods that make use of finite computational resources are employed to mimic the processes. Although molecular dynamics (MD), which calculates the atomistic behavior of molecules in a system, is a popular method for this purpose, it is frequently hampered by severe time restrictions \cite{md:1}. For example, millisecond-long calculations are still a challenge for MD  \cite{millisecond:1}, whereas the typical time scale for cell signaling, such as the mitogen-activated protein kinase (MAPK) cascade, which is thought to regulate transcription and tumor formation ranges from minutes to hours \cite{mapk:1}. Alternatively, network biology uses dynamic systems, i.e., conventional differential equations, to describe cell signaling pathways on the reaction networks between chemical species \cite{systemsbiology:1}. Although such an approximation drastically reduces the computational workload, network biology neglects spatial distribution, which compromises its real application as the diffusion rate significantly decreases in an extremely dense molecular environment in a phenomenon known as molecular crowding \cite{crowding:1}\cite{crowding:2}. In addition to the two methods above there is the lattice-based method, which is able to resolve the above-mentioned issues. In its early days, the lattice-based method employed diffusion-reaction or master equations to formulate the underlying physics and chemistry of a cell in explicitly discretized space and time \cite{latticeboltzman:1} \cite{cellsim:1} \cite{smartcell:1} \cite{latticemicrobe:1}. 

Diffusion and reactions are the two most fundamental processes in a cell, and partial differential equations (PDEs) are insufficient to incorporate inherently discrete phenomena, such as molecular crowding. To overcome this shortcoming, fully discrete mechanics in a lattice with molecular resolution is preferred . This approach explicitly discretizes space with voxels that can be exclusively occupied by a single molecule. The diffusion process then is represented by molecules hopping around the voxels. A chemical reaction takes place when reactant molecules collide, and the reaction is treated in a fully discrete and stochastic manner. To be stochastically correct, it is desirable that voxels be as isotropic and equidistant as possible. To meet such a requirement, Spatiocyte  \cite{spatiocyte:1} was created as part of the E-Cell System \cite{ecell:1}, the aim of which was to offer a modeling and simulation environment for biochemical and genetic processes. Spatiocyte employs random walks on a hexagonal close-packed lattice and a chemical master-equation based on Monte Carlo methods for chemical reactions. 

Because parallel computation has become an indispensable tool in the scientific community, we developed a new program called "pSpatiocyte." pSpatiocyte is a parallel implement of Spatiocyte and has been written from scratch to optimize its use in parallel computers. The foremost goal of pSpatiocyte is to realize faster and larger cell simulations particularly on a leadership supercomputer. Several computer programs that can simulate reaction-diffusion of molecules in intracellular space at the particle level have been developed, for example, some using discrete particles in continuous space \cite{mcell:1}  some others using discretized lattice space \cite{smoldyn:1} \cite{gridcell:1} \cite{stochsim:1} \cite{cellpp:1} \cite{cellsim:2} \cite{cellularphysiology:1} \cite{readdy:1}. To our knowledge, the work presented here is the first to report successful efficient simulations of particle reaction-diffusion systems with excluded volume effect to explicitly represent crowded nature of intracellular media to achieve scalability over a half million of cores on a distributed memory architecture. We present herein a parallel computational method that can perform calculations for a hexagonal close-packed lattice. We focus on the numerical implementation adopted by the method and show computational results with validations and estimations. We present our conclusion in the last section.


\section{Numerical Implementation}
Cell signaling starts from the binding of a signaling molecule to a membrane receptor and the transmission of the signaling molecule to the cytoplasm. The signaling molecule finally reaches the nucleus after a myriad of spatial translations and molecular transformations. Those processes can be mathematically formulated as diffusion-reaction equations from a coarse-grained point of view. The corresponding PDEs model microscopic molecular behavior as diffusion and reaction processes on a macro scale and are a basis for simulating cell signaling. To solve PDEs on digital computers under specific initial and boundary conditions, researchers usually use the finite difference method based on the Cartesian coordinate system. However, it was recently found that the diffusion rates varied both locally and temporally depending on the circumstances, which makes conventional PDEs with constant coefficients inapplicable. 

Fully discrete algorithms are expected to overcome these shortcomings. Monte Carlo simulations model diffusion and reaction processes as a random walk of molecules and collisions between molecules. They also introduce a voxel as the fundamental element, which can be recognized as a box that holds a molecule. A cell is modeled as a collection of several groups of voxels called compartments. Each molecule, e.g., protein, should reside in one voxel, and no voxel is allowed to contain more than one molecule at a time. Each molecule randomly moves to an adjacent voxel, and a reaction is set off once the voxel is occupied. 

Monte Carlo simulations sometimes result in stochastically incorrect calculations if the voxel arrangement is not taken into careful consideration. We therefore adopt a hexagonal close-packed lattice in terms of isotropy and equidistance. Furthermore, to overcome the severe time restriction inherent in differential equations, we introduce an event scheduler to schedule various time steps optimally. Chemical reactions are also treated in a fully discrete manner according to Gillespie's method and the Collins-Kimball approach. In addition, enabling parallel Monte Carlo simulations requires improvements in inter-process communication and sampling order. In what follows, we describe these implementations in detail. 

\subsection{Coordinate System}
The Cartesian coordinate system is frequently used in simulations regardless of the research field. It has six nearest neighbors for each node, i.e., voxel. However, the system is sometimes insufficient to suppress numerical noise inherent to a discretized system. To solve this problem, twelve of the second-nearest or eight of the third-nearest, i.e., diagonal, nodes are used. However, those higher order neighbors are located \begin{math}\sqrt{2}\end{math} or \begin{math}\sqrt{3}\end{math} times the distance from the origin compared with the nearest neighbors. In Monte Carlo simulations, assuming that the average velocity of a molecule is constant, the hopping duration should be longer as the distance increases. As a result, two or three intervals should be managed to incorporate such higher order neighbors. In addition, it is difficult to determine the equally distributed probabilities between distances consistently. Moreover, enumeration of the angle is not an easy task.

\begin{figure}
\centering
\includegraphics[width=2.8in]{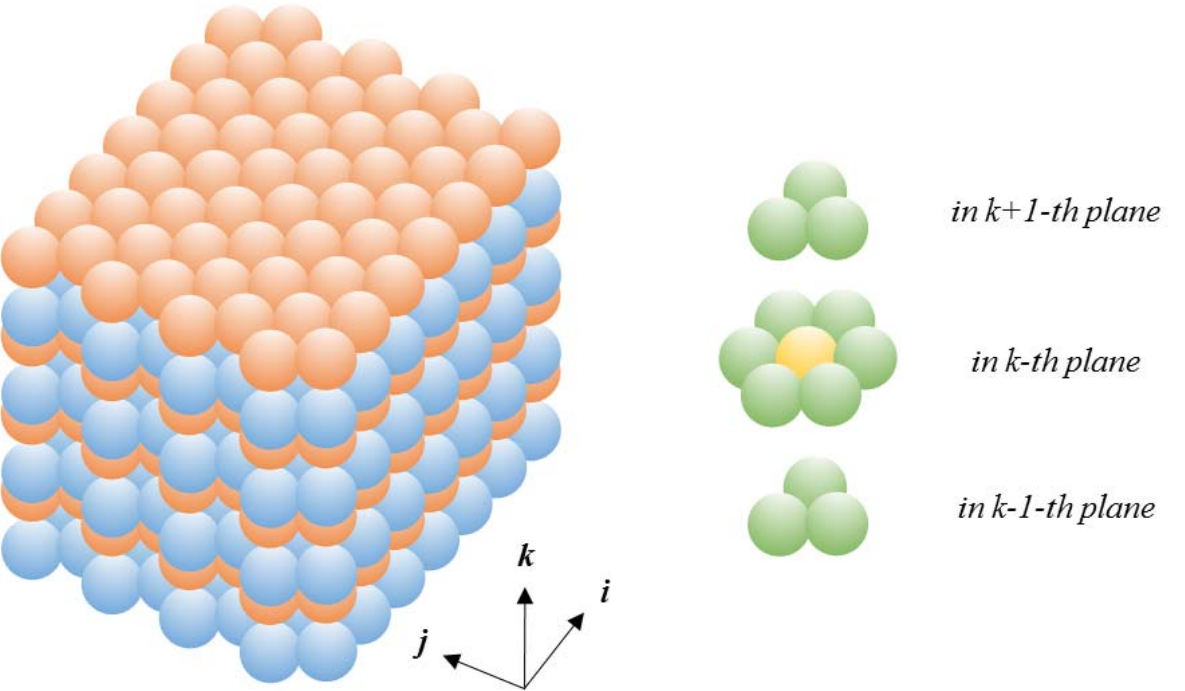}
\caption{Schematic view of an HCP lattice (left). Even and odd layers are shown in different colors. A portion of the lattice (right). Each voxel (yellow) is surrounded by twelve adjacent voxels (green). }
\label{fig:hcp}
\end{figure}

In contrast, a hexagonal close-packed (HCP) lattice has twelve nearest neighbors, twice as many as the Cartesian coordinate system (Figure \ref{fig:hcp}). From the viewpoint of spatial isotropy, adjacent neighbors forming Platonic solids are the most desirable. However, Platonic solids except hexahedrons cannot fill a three-dimensional space completely. Although the HCP lattice does not form a Platonic solid whereas the Cartesian coordinate system forms a hexahedron, the neighbors of the HCP lattice are much more densely and uniformly arranged than those of the Cartesian coordinate system. Furthermore, it has been proven that no regular lattice has more densely packed adjacent neighbors than the HCP lattice. The average density of the HCP lattice is 74.048\%, whereas that of the Cartesian coordinate system is 52.359\%. Indeed, T. C. Hales proved in 1998 that the HCP lattice is the most closely packed lattice \cite{kepler:1}.

\begin{figure}
\centering
\includegraphics[width=2.8in]{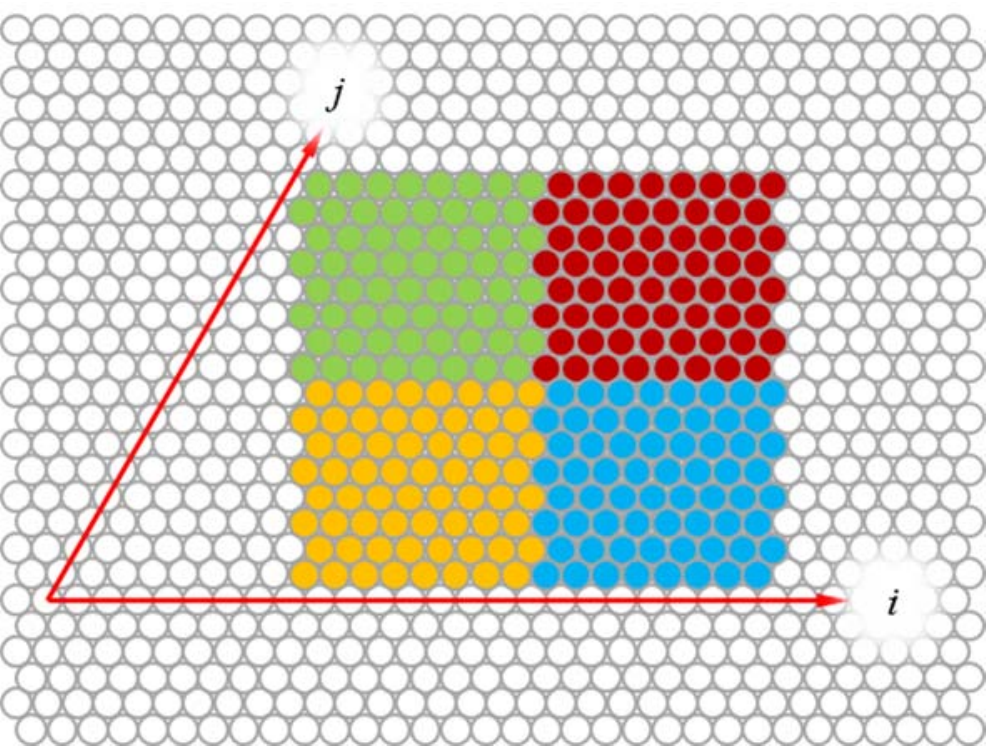}
\caption{Two-dimensional slice of a coordinate system based on a unit cell. A rectangular region mapped along angled axes i and j is recognized as a parallelogram. In addition, a tilted third axis (not shown) further modifies the boundaries of the region. Assuming domain decomposition of the region gives four parts, wedge-shaped margins complicate assignments to each three-dimensional array.}
\label{fig:unitcell}
\end{figure}

Generally, the number of neighbors is a prime concern when obtaining clean results in stochastic simulations. We therefore expect that the HCP lattice should serve as an ideal coordinate for Monte Carlo simulations \cite{spatiocyte:1}. Although the HCP lattice is a regular grid, some considerations are needed prior to the introduction of the coordinate axes. In crystallography, a coordinate system based on a unit cell is usually used, but it is far less convenient for three-dimensional simulations on digital computers (Figure \ref{fig:unitcell}). Because the unit cell of the HCP lattice is markedly squished, it is difficult to identify the coordinate of a specific neighbor in three-dimensional space. To worsen matters, unnecessarily wide margins are sometimes required to house entire computational objects. In fact, even a data structure without a coordinate system is possible. 
For example, a one-dimensional chain of voxels, such as an unstructured grid, may work. In this case, each voxel should have pointers to its neighbors. However, indirect memory access to load the data sometimes adversely impacts efficiency. In addition, an unstructured grid could still be inconvenient for a transformation, such as decomposition or merger, of arbitrary parts of an object. 

\begin{figure}
\centering
\includegraphics[width=3.4in]{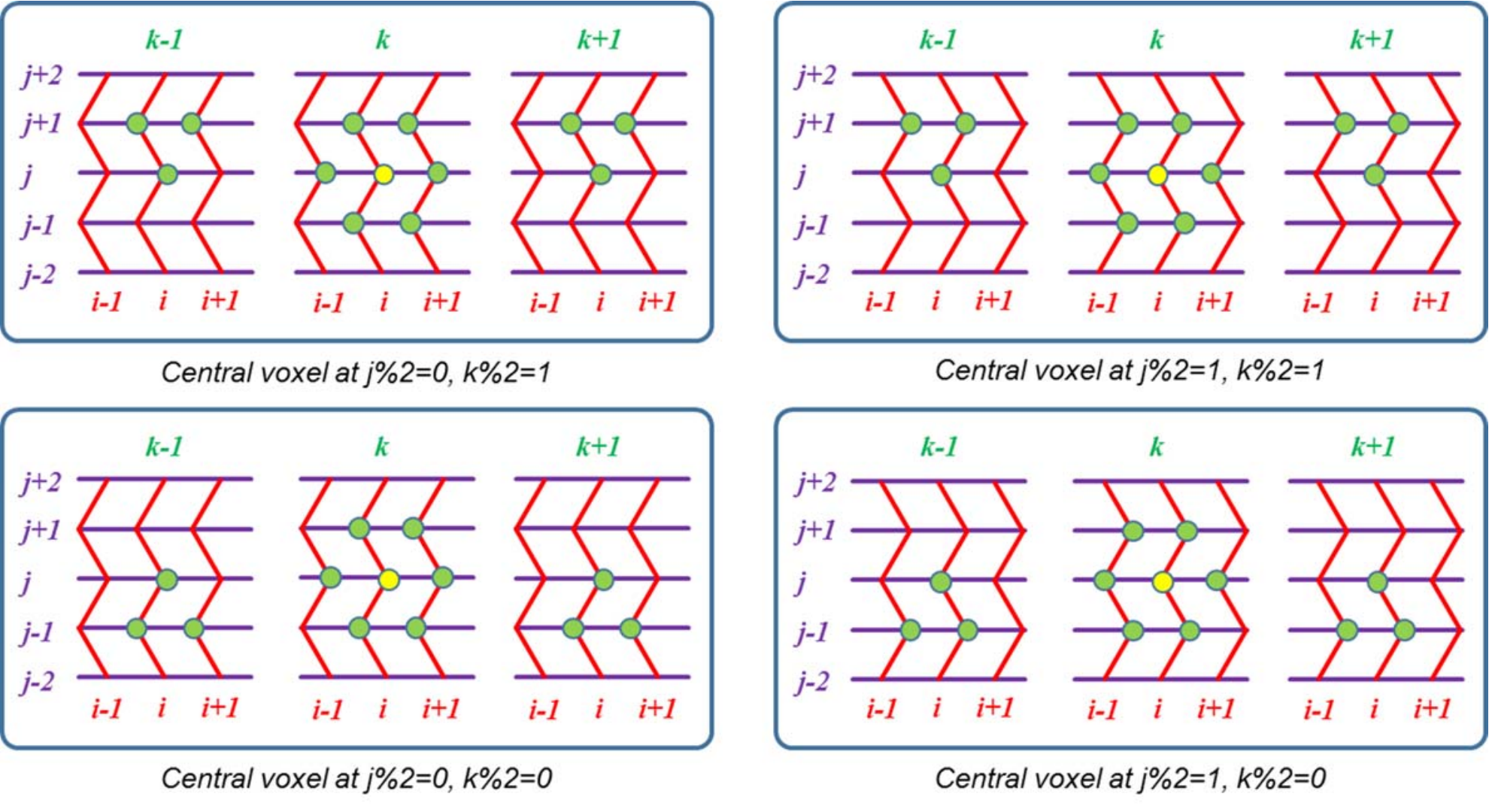}
\caption{Adjacent voxels in a twisted Cartesian coordinate system for the even plane in k-axis (bottom). For the odd plane, a slightly different stencil should be used (top).}
\label{fig:twist}
\end{figure}

To overcome these difficulties, we propose a unique coordinate system called the twisted Cartesian coordinate system, which comprises a straight axis and two zigzag axes. Although it may seem awkward, the twisted Cartesian coordinate system actually works well when identifying nearest neighbors. We have used this coordinate system for parallelization \cite{subvolume:1}. Recently a similar device was applied for cellular automata in two dimensions \cite{twistedcartesian:1}. Figure \ref{fig:twist} illustrates how a specific neighbor is identified. One of two procedures should be correctly used according to the even or odd layers in the k- and j-axes. The twisted Cartesian coordinate system can be readily mapped onto a conventional Cartesian coordinate system without almost any modification. Further, its similarity to ordinal Cartesian coordinate systems gives programmers great relief and ease. Therefore, we expected that the twisted Cartesian coordinate system be suitable to define multi-domain computational objects in an HCP lattice. In practical domain decomposition, we always restrict the number of voxels in each axis to an even number for simplicity. Readers should note that the restriction is not substantial and that an odd number of voxels is also possible theoretically. However, such a generalization requires exhaustingly complicated considerations of the boundaries and is cumbersome. Another limitation for practical use is the lack of descriptions of flat boundaries along the zigzag axes. However, this constraint is negligible for molecular scale applications.

Recently, we also found the way to apply the twisted Cartesian coordinate for face-centered cubic (FCC) which has the same density as HCP and is closer to spherical symmetry than HCP, See figures in Appendix 1 and 2 for detail.

\subsection{Event Scheduler}
In a typical numerical simulation, forward and backward finite differences are frequently employed for integration in time, which provides a straightforward and suitable approach to problems formulated with PDEs. In the course of integration, the time interval used should normally be as large as possible to fulfill stability conditions. Nevertheless, the finite difference method intrinsically manages a single time interval throughout the calculation although a number of characteristic time intervals likely exist in the problem. Eventually, the smallest characteristic time interval that fulfills the stability conditions is chosen in practice. However, stability conditions frequently impose severe limitations on the calculation of physically interesting phenomena. To alleviate such limitations, which are apt to occur in mathematically stiff problems, the operator splitting technique is usually incorporated, in which difference operators and variables are artificially split and separately treated in their numerically stable range. 

In contrast to PDEs, Monte Carlo methods can naturally realize multi-time scale procedures. Suppose molecules of different chemical species reside in some voxels. As seen in the next section, each chemical species has its own characteristic time to hop into one of the neighboring voxels, which is dependent on its weight. A straightforward way to realize random walking is to choose a molecule of a specific chemical species randomly and let it hop from one voxel to another. Note that molecules of the same species can be processed asynchronously, whereas molecules of different species can be processed only in chronological order. Regarding reactions, no matter how complex a reaction, it can be decomposed into a series of elementary reactions. Generally, each elementary reaction is classified into two types: first-order, in which a single molecule spontaneously changes its species or decays into multiple molecules, and second-order, in which two molecules collide and then transform into another species or coalesce to form a single molecule. One should keep in mind that each elementary reaction has a specific time interval similar to multi-species diffusion. 

In our program, a first-order reaction is processed every diffusion event whereas a second-order reaction is treated as an independent event. Therefore, the event scheduler should keep track of diffusion and second-order reaction events. Our scheduler is written in C++ language and is programmed to arrange multiple events correctly and quickly supply the latest event at any moment \cite{eventscheduler:1}. In practice, one should register the time intervals of all diffusion events as the first time. The event scheduler interprets events as a thread and arranges them in chronological order. Note that only a master process owns a scheduler. In the course of calculation, the master process determines the event to be taken every moment and distributes information to the worker processes. Workers interpret the received information and perform accordingly. 

\subsection{Diffusion}
As we mentioned above, in Monte Carlo simulations, the diffusion process is interpreted as numerous consecutive steps of a randomly walking molecule. The hopping frequency and distance should be consistent with the diffusion coefficients of PDEs on the macro scale. The following equation assures the consistency
\begin{equation} \lambda^2 = 6Dt. \label{eq:msd} \end{equation}
This equation is called the Einstein-Smoluchowski equation, which is named after the two prominent pioneers who derived it in the early twentieth century. In Appendix 3, we reproduce an alternative derivation of this equation from the point of stochastic differential equation view. It is also the simplest example of a general fluctuation-dissipation theorem. Physically, $\lambda$ can be understood as the mean distance a molecule can travel in a given time interval. Taking $\lambda$ as the voxel diameter, $t$ can be interpreted as the mean time $\tau_d$ each molecule takes to move in adjacent voxel. 
\begin{equation} \tau_d = {2 r_v^2 \over 3 D}, \end{equation}
where $r_v$ denotes voxel radius. Eventually,  $\tau_d$ is equal to the time interval between diffusion events. It should be noted that the time interval derived herein differs according to chemical species. Therefore, the diffusion process should be processed for each species.

To understand what is taking place during program execution, suppose an event scheduler returns a diffusion event of the $i$-th species. Then, an adjacent voxel is randomly chosen for each molecule of species $i$, and each molecule moves to the chosen voxel only if the voxel is vacant. This movement fails if the voxel is already occupied by another molecule or if the voxel is in a different compartment, e.g., membrane. Molecules are processed one by one in dictionary order. The processing order of molecules matters theoretically, but practically the difference is always negligibly small. Furthermore, this asynchronous property is indispensable for the parallel processing of stochastic calculations. Finally, the next diffusion time is updated by adding $\tau_d$ to the current time.

\subsection{Reactions}
Normally, no matter how complicated a chemical reaction, it can be generally decomposed into elementary reactions that obey the simple law of mass action. Recall that an elementary reaction is set off by collisions between reactants. Because the probability of more than three molecules meeting somewhere at a time is thought to be infinitesimally small, considering only two reactants often suffices. A single reactant may spontaneously transform into another molecule or decay into multiple molecules after a specific duration. In contrast, two reactants may coalesce into one or decay into multiple molecules. These reactions are called first- and second-order reactions, respectively. The first-order reaction occurs independent of a collision, whereas the second-order reaction is dependent. Consequently, the second-order reaction is a diffusion-influenced process. Therefore, we should realize these reactions numerically by using different procedures according to their underlying physics. 

Note that the number of products could be more than three and is limited to less than twelve, $i.e.$, the number of nearest neighbors, in our method. In an extreme situation, eleven neighbors could be already occupied and no reaction would take place. This condition is highly unlikely and can be avoided by introducing a multi-stage reaction that decomposes a multi-product reaction into a cascade of bi-product reactions. Another factor to consider is the voxel size. When the voxel is too large compared with the molecular size, exclusive volume is overestimated. Consequently, the reaction rate would be underestimated. 

\subsubsection{First-Order Reaction}
We employed Gillespie's algorithm with slight modification for a given duration to realize a first-order reaction numerically \cite{gillespie:1}. The algorithm can reproduce a stochastically correct law of mass action. In this method, called Gillespie's direct method, a species dependency table is prepared from a directed graph of species. Suppose the following chemical formula,  
\begin{eqnarray*} 
\mathrm{A} &\overset{k}{\rightarrow}& \mathrm{B} + \mathrm{C} + \dots .
\end{eqnarray*}
The propensity is calculated for each reaction with the concentration of reactant species as follows: 
\begin{equation}
a = k c_A
\end{equation}
where $a$ and $c_A$ are propensity and the concentration of species A, respectively. In every diffusion event, the propensities of the provided first-order reactions are calculated, and the probability $P$ that any first-order reaction is fired can be calculated, as follows: 
\begin{equation} 
-\ln P = ( t_{current} - t_{previous} ) a_0, \quad \left( c.f. \ a_0 = \sum^m_{i=1} a_i \right)
\end{equation}
where $a_0$, $a_i$, $m$, $t_{current}$, and $t_{previous}$ denote total propensity, propensity of the $i$-th first-order reaction, the number of provided first-order reactions, current time, and time of the previous diffusion event, respectively. 
If a random number drawn from [0;1] is larger than $P$, one of the first-order reactions is fired. We determine the $u$-th first-order reaction that satisfies the following condition: 
\begin{equation}
\sum^{u-1}_{i=1} a_i < a_0 R \le \sum^{u}_{i=1} a_i,
\end{equation}
where $R$ denotes a random number drawn from [0;1]. After the selected $u$-th first-order reaction is fired, species related to the fired reaction are processed. A reactant is eliminated from the voxel where it has resided and products are created. In the case of a single product, the created product is located in the same voxel where the reactant had resided. If there are multiple products, the program tries to find adjacent vacant voxels. The reaction is a success if sufficient voxels are found; otherwise, it is a failure. The concentrations of the eliminated reactant and product(s) created by the reaction are updated. In addition, propensities depending on the updated species are recalculated by consulting the dependency table. This method can reduce computational operations for updating, which requires load and store operations between the register and the main memory, and are very time-consuming, thus leading to enhanced system performance.

\subsubsection{Second-Order Reaction}
To realize a diffusion-influenced reaction, we employ the Collins-Kimball approach \cite{Collins:1} in a discretized fashion. Suppose a second-order reaction between species A and B, 
\begin{eqnarray*} 
\mathrm{A} + \mathrm{B} &\overset{k}{\rightarrow}& \mathrm{C} + \mathrm{D} + \dots . 
\end{eqnarray*}
A molecule of species A collides somewhere with a molecule of species B. The probability $p_{AB}$ of setting off a reaction is expressed as: 
\begin{equation} p_{AB} = { k_{AB} \over 6\sqrt{2} (D_A + D_B) r_v }, \end{equation}
after some manipulation \cite{spatiocyte:1}. Collisions may occur in the midst of the diffusion process in our program. Eventually, second-order reactions are executed as part of the diffusion event with the above probability. Suppose a molecule of species $A$ moves to one of its adjacent voxels. If that voxel is vacant, the diffusion succeeds; otherwise, collision occurs. A molecule in a target voxel reacts with species $A$ only if the chemical formula between the molecules is registered in the reaction table. 

\subsection{Parallel Programming}
pSpatiocyte is written in C++ language \cite{cplus:1} in accordance with object-oriented programming and is markedly reliant on the standard template library (STL) \cite{cplus:2}. It also borrows a pseudo-random number generator based on the Merssene-twister algorithm \cite{mersenne:1}. In addition, C++ bindings of the message passing interface (MPI) version 2.2 are used to enable inter-process communication \cite{mpi:1}. Thread parallelism through the OpenMP interface version 3.0 \cite{openmp:1} is also introduced to enable hybrid parallelism. Domain decomposition and a master-worker model are employed as the fundamental framework of our program. 

The most difficult problem inherent to parallel stochastic simulations, such as the Monte Carlo method, is maintaining consistency on each domain boundary. Processes should be always aware of ghost voxels residing in adjacent processes. Suppose a molecule moves beyond the boundary. The vacancy of the voxel on the opposite side should be examined. If the voxel is vacant, it must be blocked immediately to enable mutual exclusion between involved processes. Because many molecules are moving around simultaneously on both sides, maintaining the consistency of every ghost voxel is a formidable task. Busy waiting is a possible straightforward solution to mutual exclusion. Unfortunately, it works only for two alternating processes and prevents a process from reading the voxel consecutively. A standard solution to randomly accessing two parties is known as Peterson's algorithm \cite{peterson:1}. Because multiple threads may access a voxel in a diagonal ghost, the algorithm should be improved for application to a multi-party environment. To make matters worse, inter-process communication is required every time a molecule moves beyond the boundary. As a result, many MPI calls carrying short messages would only waste latency times.

Alternatively, a rolling back strategy can be used. However, this approach cancels the movement sequence. Such a sequence tends to bifurcate many times, propagates upstream beyond the boundary, and rapidly piles up until it becomes difficult to track. Generally compromising local modification and global consistency is a difficult problem. Therefore, we loosened the complete consistency somewhat to enable parallel calculations. Suppose an HCP lattice on each process forms a cube projected on a Cartesian coordinate system. We further divide the cube into octal cubes called sub-volumes. One can easily find that there are $8!=40,320$ ways to reorder these sub-volumes. 

\begin{figure}
\centering
\includegraphics[width=2.8in]{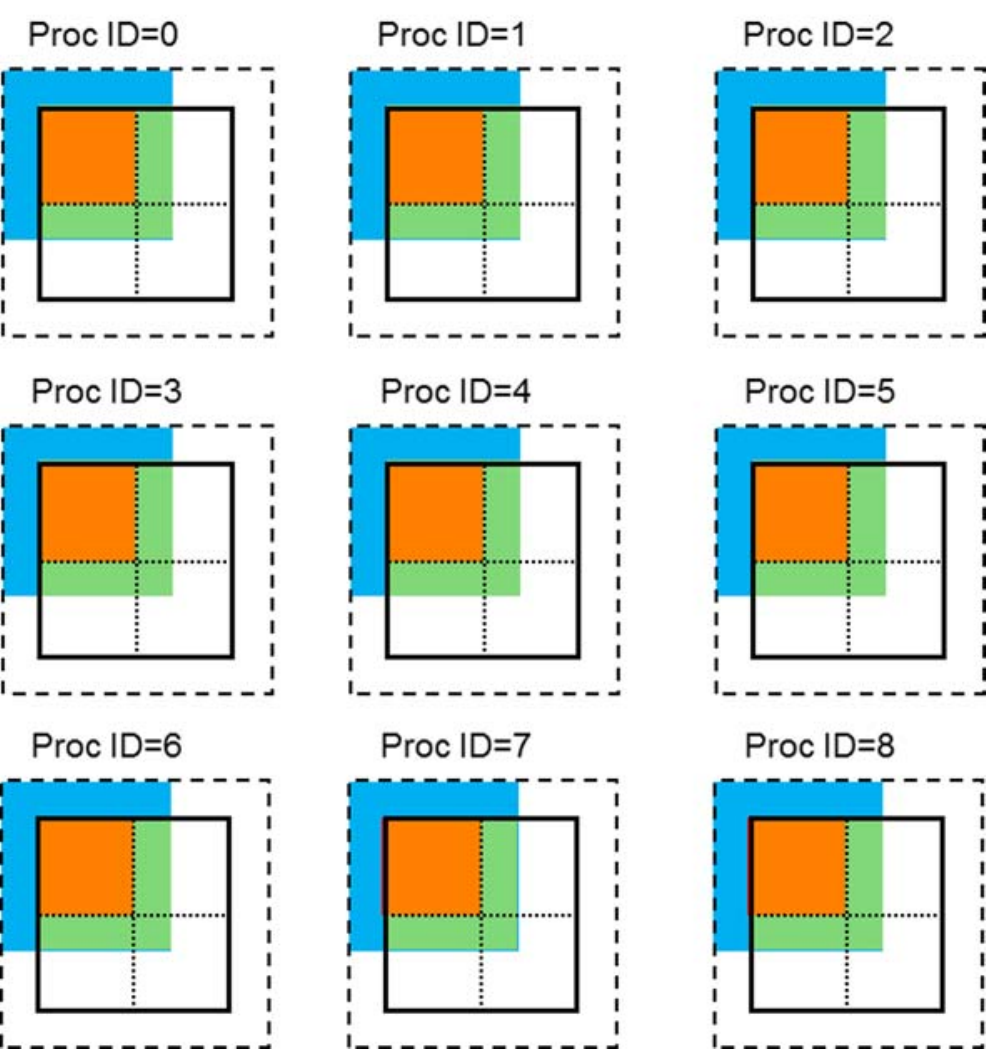}
\caption{Sub-volumes in the case of nine processes arranged in two dimensions. Thick and dashed lines represent the boundary and ghost regions of computational domains and the respective ghosts in each process. For example, the ghosts (blue and green) of every top-left sub-volume (orange) are located remotely from each other. Therefore, interference between them will never occur as long as these sub-volumes are processed simultaneously. The blue regions are obtained from adjacent nodes through inter-process communications. Four sub-volumes in each process are processed in random order.}
\label{fig:subvolume}
\end{figure}

Overall, the baseline procedure of our program is as follows: 
\begin{enumerate}
    \item Identify a sub-volume to be processed randomly,
    \item Obtain data in ghost regions from adjacent processes,
    \item Do Monte Carlo calculations,
    \item Move to the next time step if all sub-volumes are processed; otherwise jump back to 1.
\end{enumerate}
Each process calculates the same sub-volume at the same time. Sub-volumes and ghost regions are illustrated in Figure \ref{fig:subvolume}. 
Note that each sub-volume in the calculation is located remotely at any time, and the ghost voxels remain intact by adjacent processes. Although mutual exclusion is naturally realized by this method, inter-process communication should be executed eight times for each step. However, this method of mutual exclusion does not necessarily mean that eightfold communication times are required, because the number of voxels to be sent or received is also reduced in proportion to the surface area. We have confirmed the effectiveness of our method in at least a few thousand processes \cite{subvolume:1}.

\begin{figure}
\centering
\includegraphics[width=3.2in]{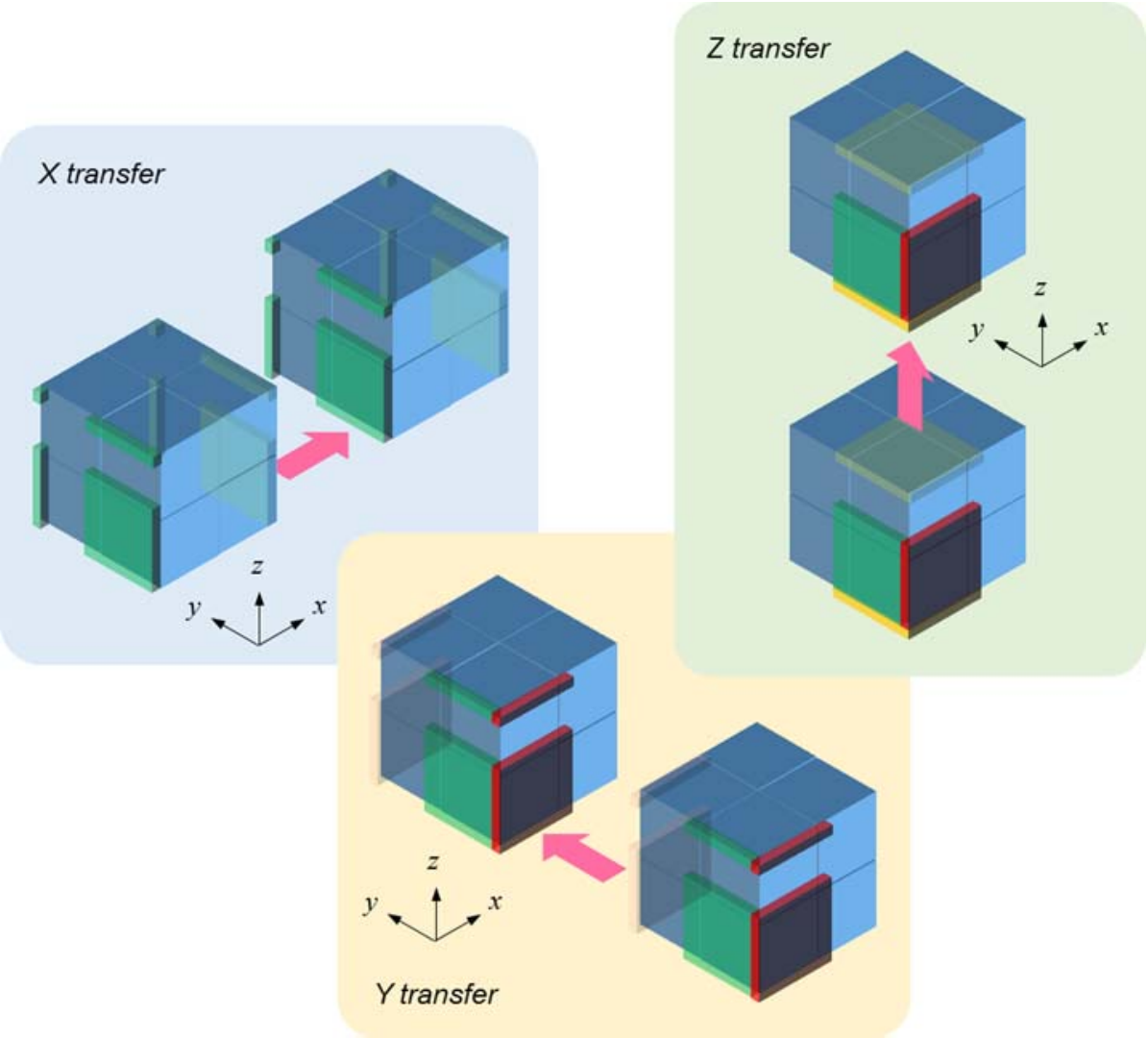}
\caption{Tri-stage lock-step data transfer. Halo of the lower front eighth of cube is tranmitted by consecutive transfer X through Z.}
\label{fig:triphase-comm}
\end{figure}

Another problem common to lattice-based parallel calculations is diagonal communication. Some ghost voxels sometimes locate in diagonally adjacent processes. In that case, one process should communicate with the other twenty-six processes surrounding it. Generally, latency has an impact on short communication. In addition, contention between requests tends to occur due to the limited number or bandwidth of the communication channels. From the viewpoint of strong scaling, such a communication strategy will show poor performance for a larger number of processes. To overcome such constraints, we employ interlocking tri-phase communication (Fig. \ref{fig:triphase-comm}), which is found in conventional literature \cite{textbook:1}. By using this method, each node suffices to communicate with six directly adjacent processes in three consecutive phases. To apply the method to our program, we further modified it so that it would fit octal sub-volume configurations. Finally, reactions have little impact on performance since they are far fewer than diffusion events.. 

\begin{figure}
\centering
\includegraphics[width=3.2in]{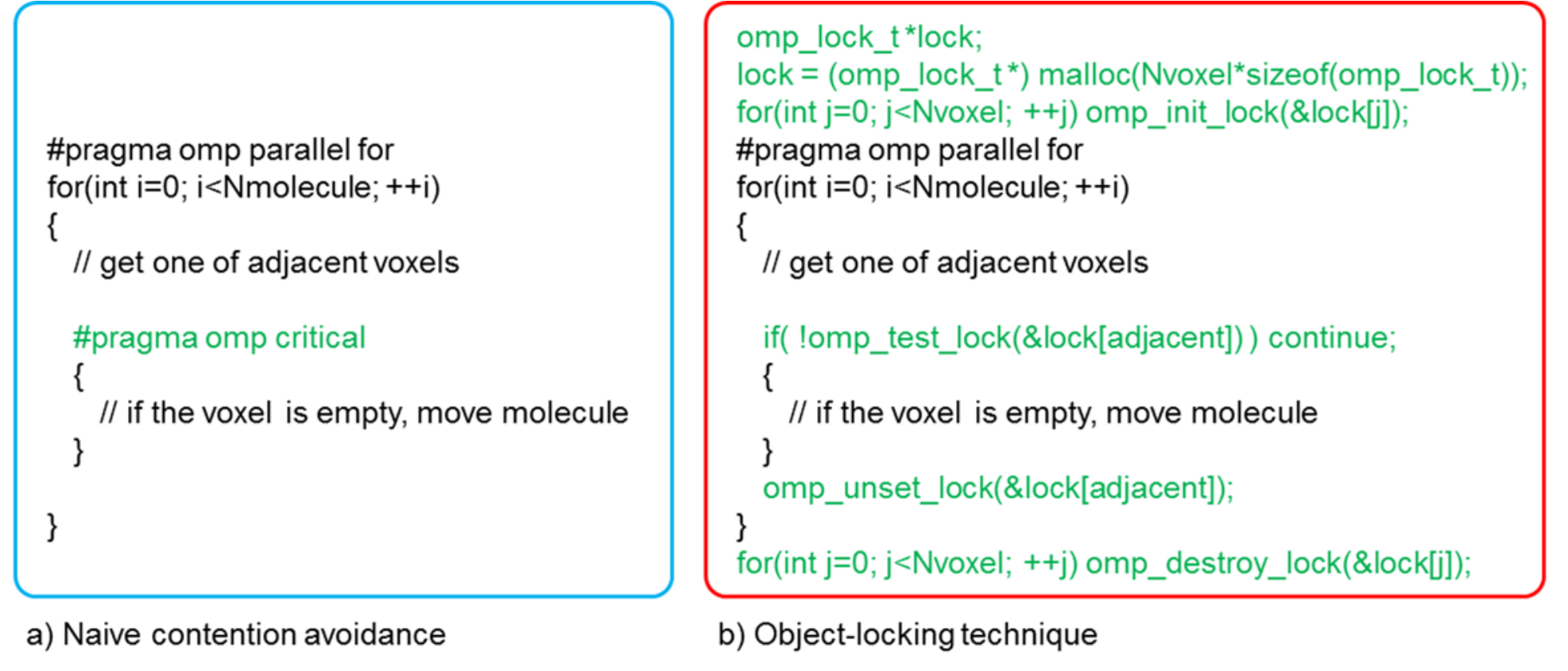}
\caption{Voxel-locking technique. a) straightforward avoidance of access contention, b) the object-locking technique by virtue of OpenMP lock variables.}
\label{fig:voxellock}
\end{figure}

As for thread level parallelization, straightforward implementation to avoid competing access to specific voxel employs critical clause from OpenMP interface. However it generally suffers poor performance even in a case of few threads since threads are apt to fall in an idle state at the entrance of critical section. To address this issue, we propose the voxel-locking technique which locks every voxel independently with OpenMP runtime routines $omp\_init\_lock$, $omp\_test\_lock$, $omp\_unset\_lock$ and $omp\_destroy\_lock$ as shown in Fig. \ref{fig:voxellock}. Although it may require millions of lock variables, each variable consumes just a small amount of byte. Therefore the impact on memory system is limited.

\section{Computational Results}
To prove the physical correctness of pSpatiocyte, we validated diffusion and reaction processes. Parallel performances were also examined across thousands of computational cores. Finally, calculation of a simplified MAPK model is shown as a preliminary result and  an example of real-world applications. All calculations discussed below were performed on the K computer
 \cite{kcomputer:1}.

\subsection{Validating Diffusion}
\begin{figure}
\centering
\includegraphics[width=2.6in]{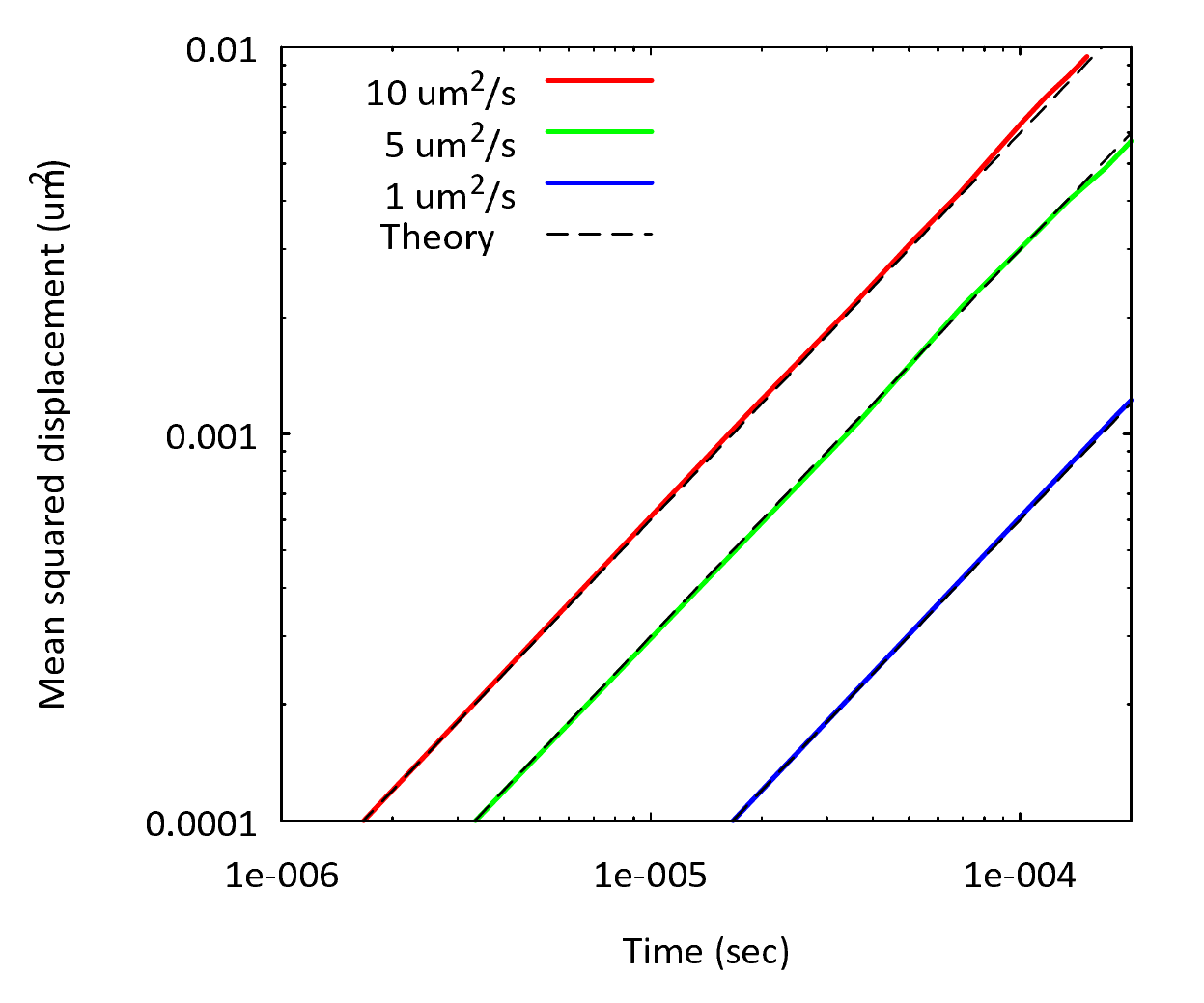}
\caption{Diffusion rates. Red, green, blue, and dashed lines represent simulated results with diffusion rates of $10$, $5$, and $1\,\mu m^2/s$ and theoretical results, respectively.}
\label{fig:diffusion}
\end{figure}

We first validated the dependence of the diffusion rates. The mean squared displacement (MSD) from the origin against time was predicted by the random walk theory as 
\begin{equation} \log \left( \lambda^2 \right) = \log t + \log \left( 6D \right), \end{equation}
which is derived from the Einstein-Smoluchowski equation. To confirm the reliability of the dependence of the diffusion rates, we monitored a single test molecule in a $960^3$ lattice with voxels measuring $2.5$ nm in radius. No other molecules existed in the lattice. Then we performed random walks repeatedly under the same initial conditions and environments aside from random number seeds. In those calculations, variance, $\sigma$ was determined from a given $D$ as shown in Appendix 3. One thousand random walks were generated. Each walk was terminated when it reached a boundary. Finally, the average locations were taken from the ensembles. Figure \ref{fig:diffusion} shows the log-log plots of the results for three diffusion rates. The slopes, the vertical distances, and the absolute values coincide with theory. Isotropy was found to have fairly good agreement at least for one thousand ensembles. 

The diffusion rates are thought to decrease under the crowding condition. To prove that pSpatiocyte is able to predict diffusion rates, we calculated MSD for three different occupancies of molecules. Fig. 8 shows the results. A significant decrease of the diffusion rate corresponding to an increase in the number of crowding molecules is clearly observed. Note that standard diffusion theory is based on the assumption that molecules are infinitesimally small and move around without collision. Finite size effects, such as volume exclusion, are not captured by the theory. 

\begin{figure}
\centering
\includegraphics[width=2.6in]{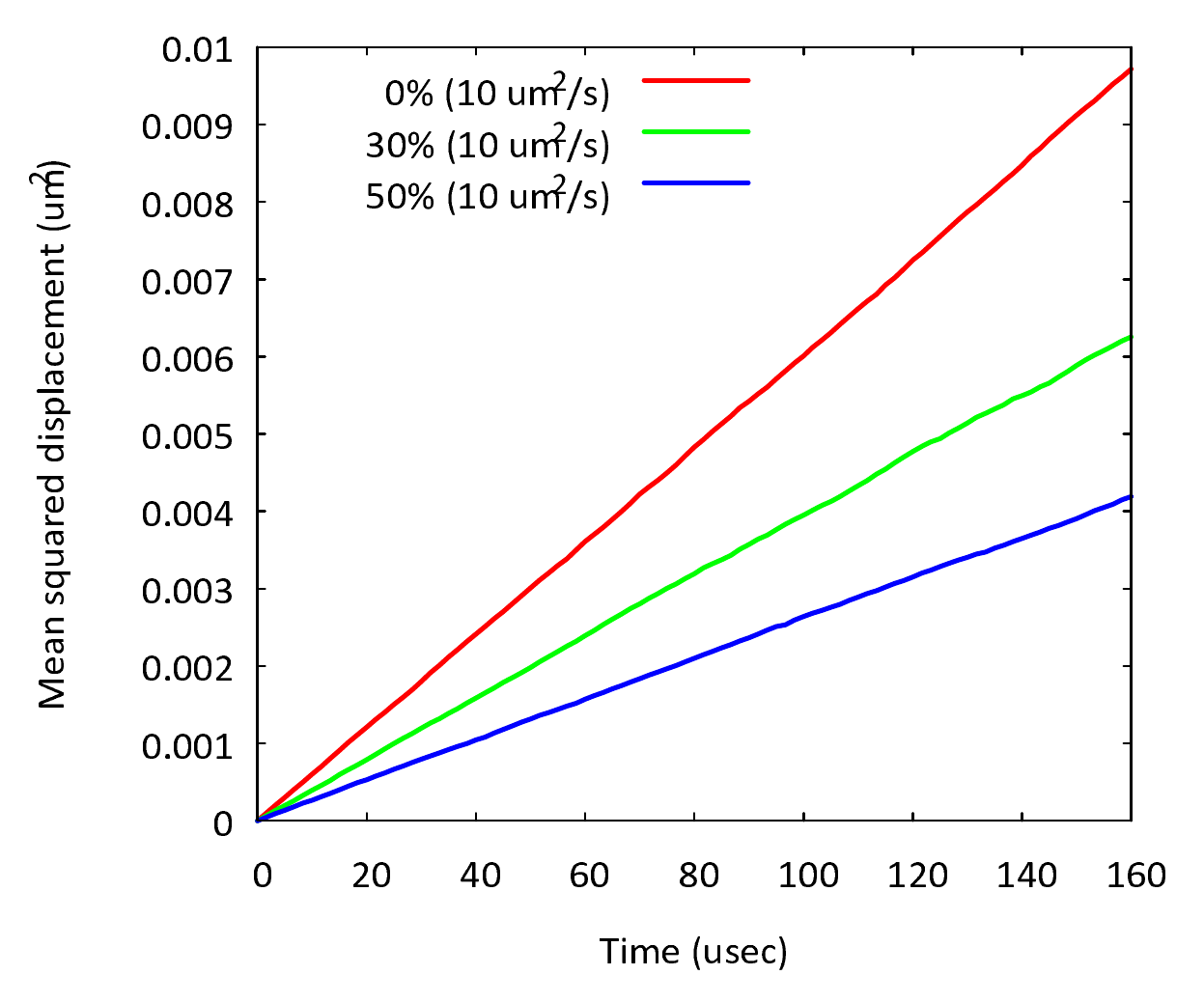}
\caption{Crowding effects. Red, green, and blue lines represent the simulated results when 0\%, 30\%, and 50\% of the lattice is occupied, respectively.}
\label{fig:crowding}
\end{figure}

\subsection{Validating Reactions}
As we mentioned earlier, the underlying physics of the first- and second-order reactions are different, thus we validated them separately using simplified irreversible and reversible reactions.

\subsubsection{Irreversible Reaction}
\begin{figure*}[ht]
\centering
\includegraphics[width=6.4in]{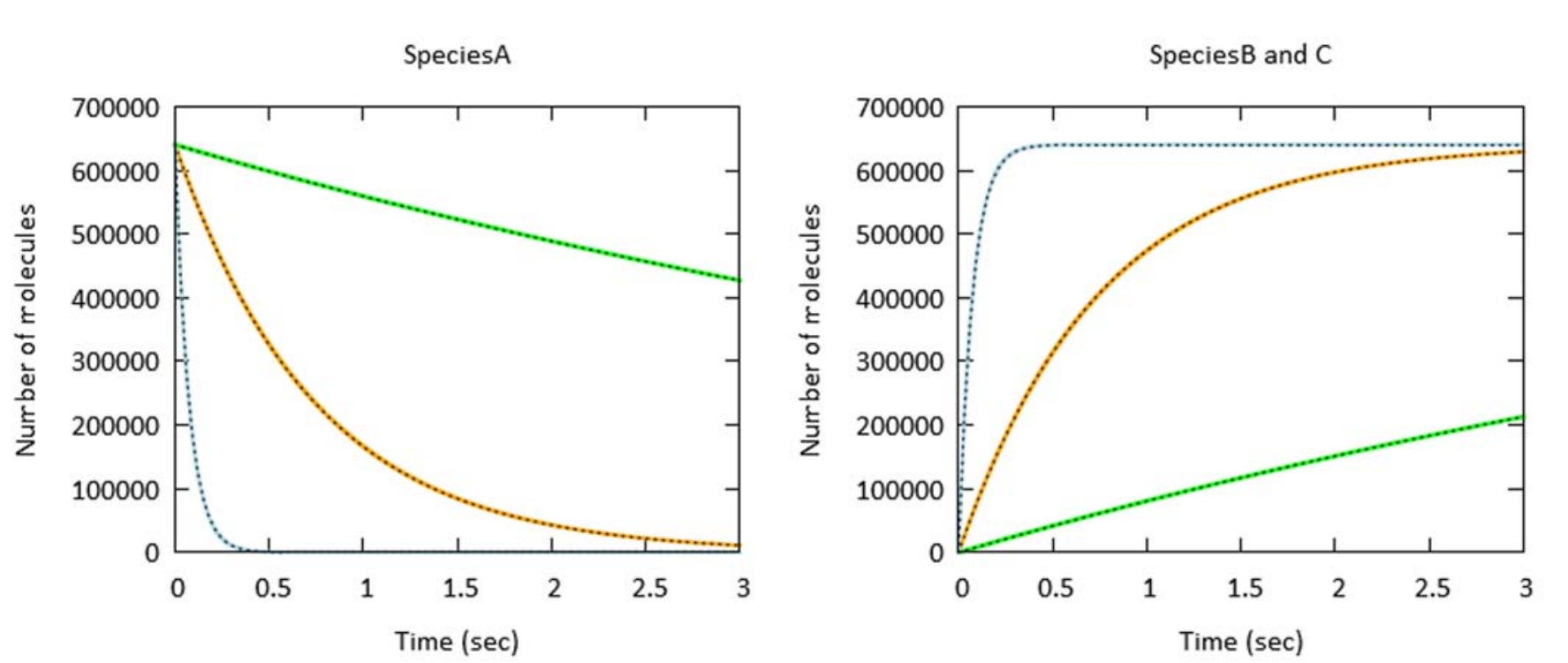}
\caption{First-order reactions. Species A (left) is divided into Species B and C (right). Green, orange, cyan, and dashed lines represent original kinetic rate (0.135), 10 times the original (1.35), 100 times the original (13.5), and theoretical results, respectively.}
\label{fig:firstorder}
\end{figure*}

A first-order reaction primarily models auto decay, such as dissociation. A reaction involving two products is typical and is written in a chemical formula as follows: 
\begin{eqnarray*}
\mathrm{A} &\overset{k}{\rightarrow}& \mathrm{B} + \mathrm{C}. 
\end{eqnarray*}
The first-order reaction is a one-way (i.e., irreversible) reaction, and its propensity depends on a single reactant. As long as the accompanying rate equation holds, the reaction can be solved analytically as follows: 
\begin{eqnarray*}
[\mathrm{A}_t] &=& [\mathrm{A}_0] \exp \left( -kt \right) \cr
[\mathrm{B}_t] &=& [\mathrm{B}_0] + [\mathrm{A}_0] \left( 1 - \exp \left( -kt \right) \right) \cr
[\mathrm{C}_t] &=& [\mathrm{C}_0] + [\mathrm{A}_0] \left( 1 - \exp \left( -kt \right) \right),
\end{eqnarray*}
where brackets and subscripts stand for species concentration and time, respectively. One should keep in mind that the rate equation holds only for low concentrations. Propensities are expected to be less than the law of mass reaction under the crowding condition because of the scarcity of free space in the surroundings. We calculated three different reaction rates within the dilute realm and compared the results with theory. The parameters used in the calculation were $k = 0.135, 1.35 \; or \; 13.5 \, s^{-1}, D = 10 \,\mu m^2/s$ and $r_v = 5 \,nm$. The initial numbers of molecules of species $A$, $B$ and $C$ were 64,000, 0 and 0, respectively, and lattice size was $960^3$. The results are shown in Figure \ref{fig:firstorder}. 
Good agreement was observed for each case, however, validation of the denser case was necessary for reliable cell signaling simulation. For the time being, the results of pSpatiocyte agree with those of our precedent program, Spatiocyte \cite{spatiocyte:1}, despite ambiguity in the probability estimation.

\subsubsection{Reversible Reaction}
\begin{figure*}[ht]
\centering
\includegraphics[width=6.4in]{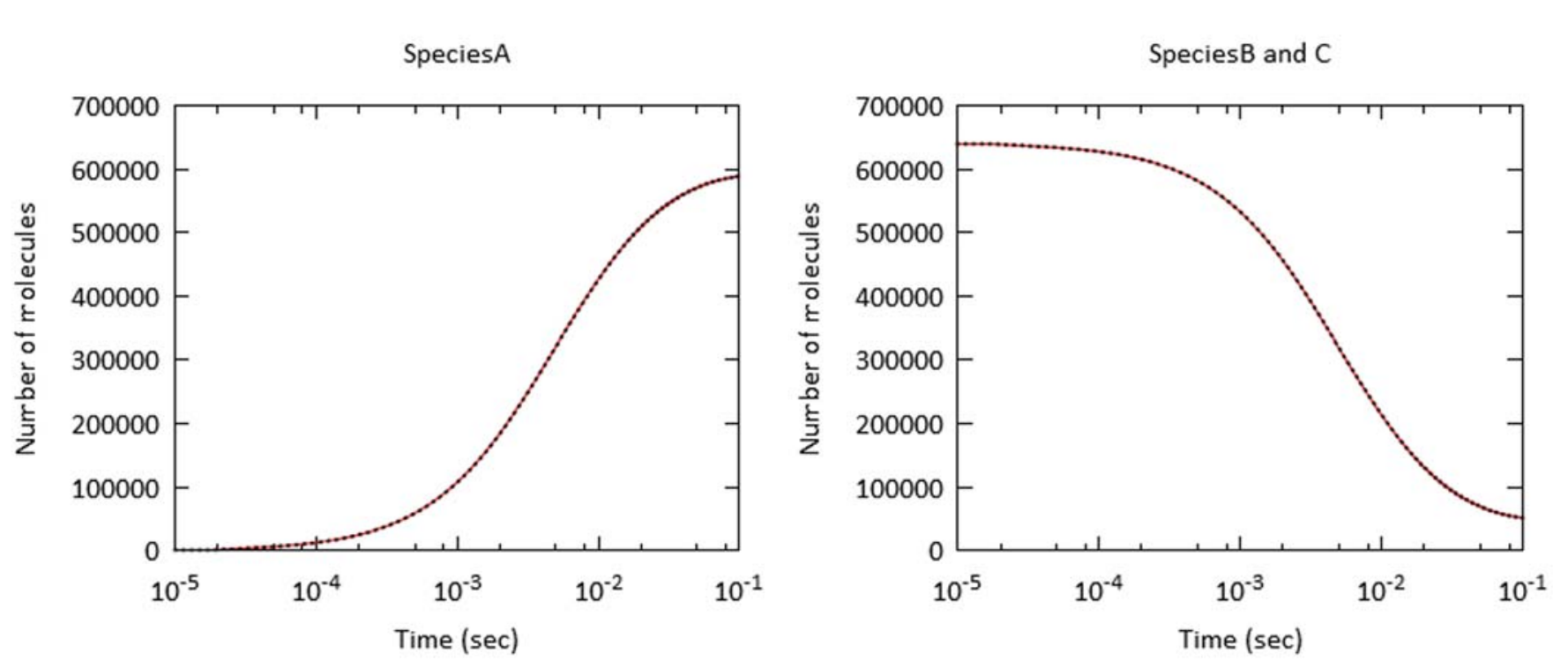}
\caption{Reversible reaction. Species A (left) is divided into Species B and C (right) by forward reaction, and simultaneously Species B and C form Species A by backward reaction. Red and dashed lines represent simulated results by pSpatiocyte and Spatiocyte, respectively.}
\label{fig:secondorder}
\end{figure*}

A second-order reaction is fundamentally diffusion-influenced because it is set off by reactants colliding. For simplicity, we considered a single product instead of bi-products as the simplest example of a reversed reaction: 
\begin{eqnarray*} 
\mathrm{B} + \mathrm{C} &\overset{k_f}{\underset{k_b}{\rightleftarrows}}& \mathrm{A}, 
\end{eqnarray*}
where $k_f$ and $k_b$ denote forward and backward reaction rates, respectively. Unfortunately, no closed-form analytical solution convenient for comparison is available even for the simple formula above. Therefore, we used our serial program, Spatiocyte \cite{spatiocyte:1}, as a measure. The parameters used were $k_f = 0.027\,nM^{-1}s^{-1}, k_b = 1.35\,s^{-1}, D = 10\,\mu m^2/s$ and $r_v = 5\,nm$. The initial numbers of molecules of species A, B and C were 0, 0 and 64,000, respectively, and the lattice size was $1000*960*1100$. 
The results are shown in Figure \ref{fig:secondorder} with the temporal axis plotted in logarithmic scale to examine changes in detail. 
Good agreement is noted between the results of the two programs. Note that the reaction is diffusion-influenced in the case of low concentration, such as this example. Therefore, the observed reaction rate was expected to be lower than that predicted by the law of mass action. In the case of high concentration, the reaction rate will increase gradually to the value the law of mass action predicts.

\subsection{Parallel Performance}
To estimate the parallel performance of our program, we compared real time from the point of view of strong and weak scaling. Strong scaling measures how fast a program is able to process specific problems by parallelization. On the other hand, weak scaling measures how large of a problem a program is able to handle without loss of speed.

For strong (i.e., intensive) scaling, three resolutions according to the voxel radius were used to measure performance. Provided that the physical dimensions of a cell remain the same, a smaller voxel would serve as a finer lattice. We denoted voxels having 10, 5, and 2.5 nm radii as coarse, intermediate, and fine lattices, respectively. The parameters and conditions were the same as those of the diffusion cases in the previous section, except for lattice size and occupancy. Occupancy was fixed at 30\% in this section, and the the whole lattice size was $512^3$(coarse), $1024^3$(intermediate) or $2048^3$(fine) except fot calculations using 663552 cores, the largest number of available cores on K computer, in which we used $512x480x540$ (coarse), $1024x960x1080$ (intermediate) or $2048x1920x2160$ (fine) lattices to make it conform with the physical configuration of nodes.

\begin{figure}
\centering
\includegraphics[width=2.8in]{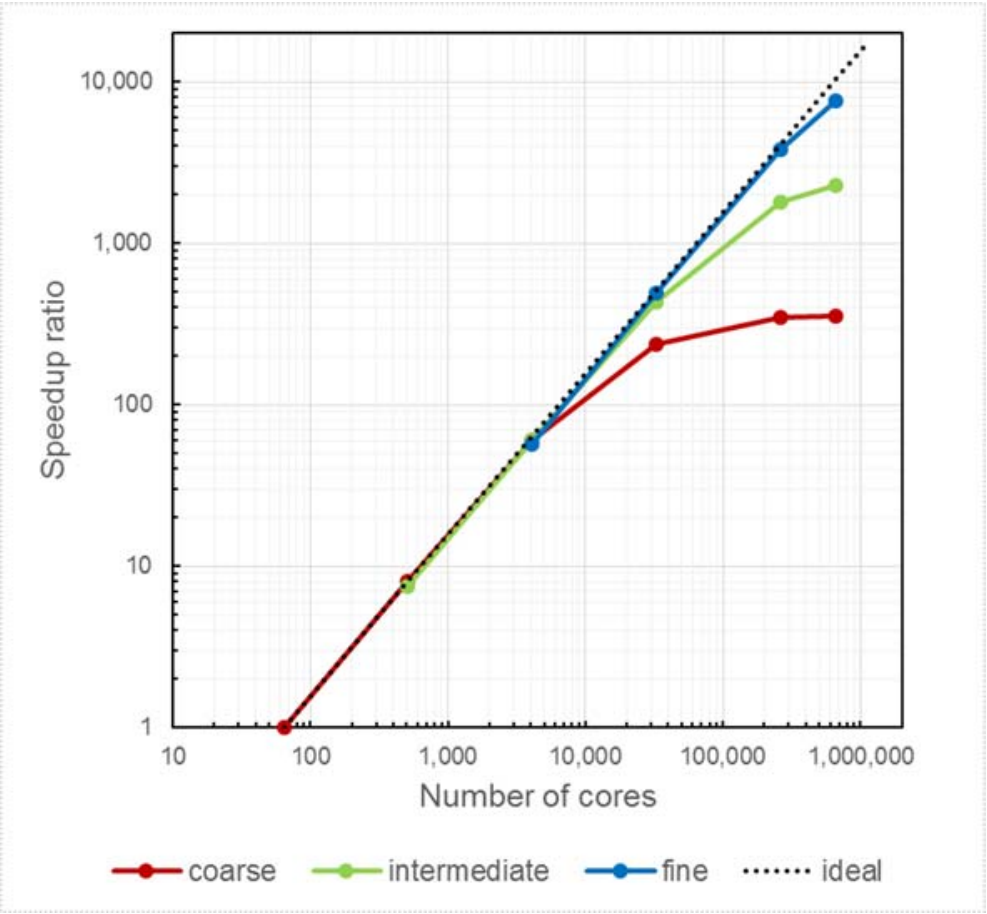}
\caption{Speedup ratios relative to 64 cores. Red, green, and blue lines represent lattices with $512^3$, $1024^3$, and $2048^3$ voxels, respectively.}
\label{fig:strong}
\end{figure}

\begin{figure}
\centering
\includegraphics[width=3.2in]{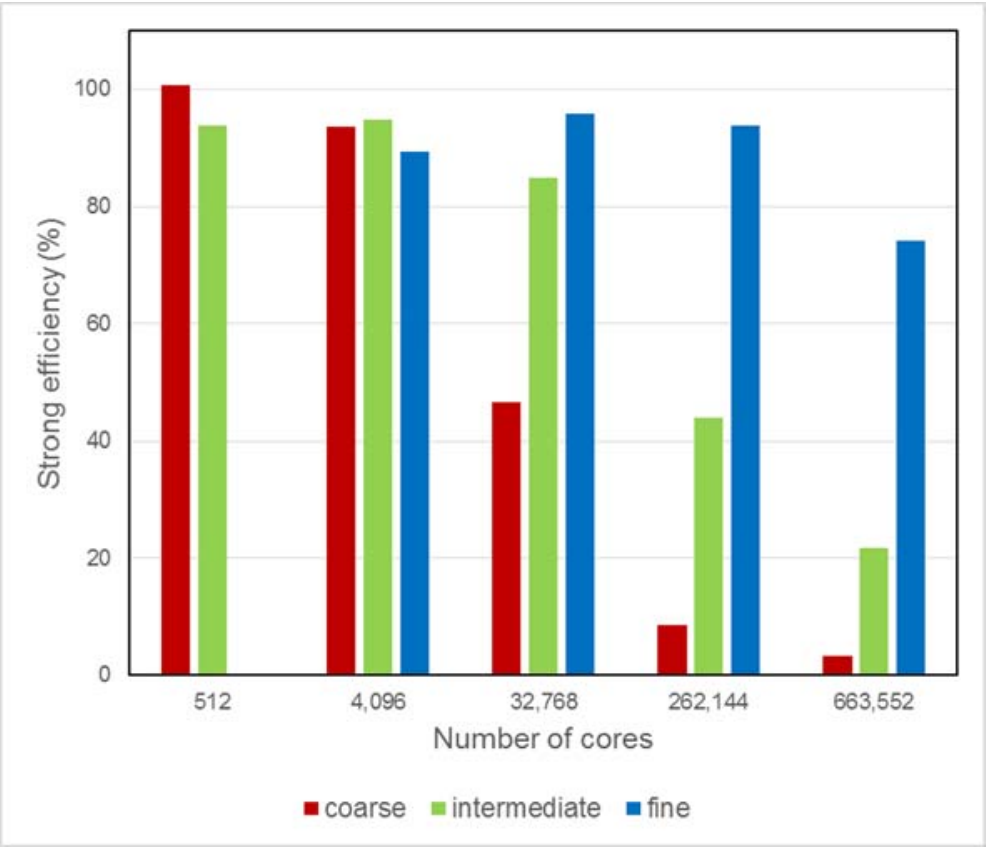}
\caption{Strong efficiency. Colors are the same as those in Fig. \ref{fig:strong}.}
\label{fig:strongefficiency}
\end{figure}

\begin{figure}
\centering
\includegraphics[width=3.2in]{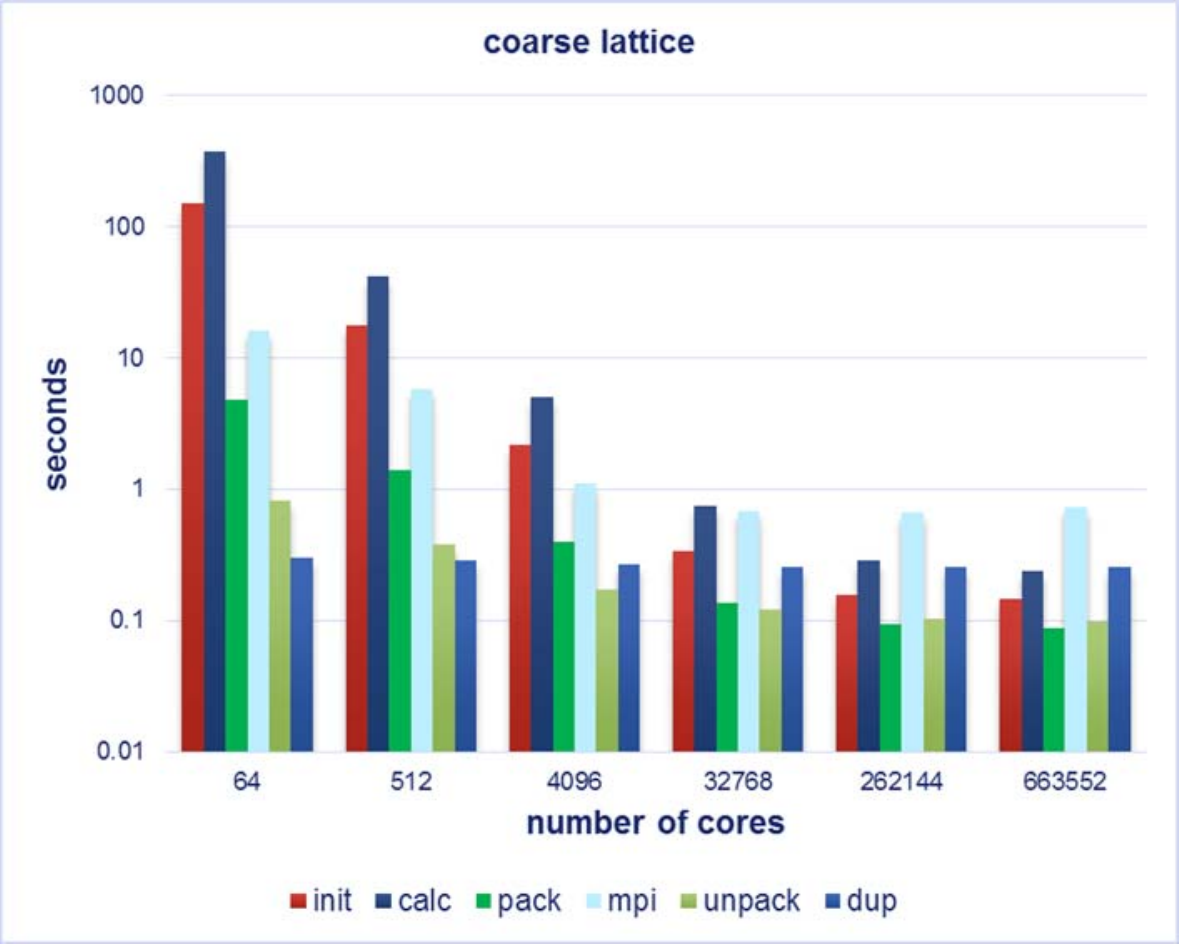}
\caption{Init and calc show times for initialization and computation. Pack and unpack are times taken for pre- and post-communication data managements, mpi is a time spent on MPI\_Sendrecv. Dup is a time taken to duplicate communicator objects.}
\label{fig:component}
\end{figure}

Notice that we examined relative speedup rates instead of bare FLOPS since a large part of the calculation was spent in integer or logical instructions. The speedups measured from the elapsed times are shown in Fig. \ref{fig:strong}. In this figure, we took the result of the coarse lattice with 64 cores as the denominator as simulation with less cores was not possible due to memory restriction. For intermediate and fine lattices, data of 64 cores were extrapolated from consumed time per voxel of coarse lattice. In addition, parallel efficiencies of strong scaling are also summarized in Fig.  \ref{fig:strongefficiency}. For the fine lattice with 663552 cores, speedup of 7686 times was achieved. Parallel (strong) efficiency of that case was 74.1\%. In contrast, the other, especially coarse, lattice shows conspicuous separation from the ideal curve. 
To identify the cause of deterioration, we closely figured out the components of elapsed time of coarse lattice as shown in Fig.  \ref{fig:component}. In this figure, we find that initialization, calculation, pack and unpack times kept decreasing at least up to 262144 cores while MPI time saturated and overwhelmed others over 32768 cores. Constant duplication time due to redundant duplication of communicator object should be eliminated by sophisticated programming in the next version. As a result, saturation of MPI time is thought of as the most significant factor to improve scaling further. However, saturation of that kind generally originates from the latency of inter-node communication which is specific to hardware or firmware.

Consequently, all we can do to improve strong scaling is to decrease calculation time as close to latency as possible. In this sense, deterioration observed here is inevitable and beyond programming effort. Lastly, by extrapolating from the discussion above, we can predict the speedup and efficiency of fine lattice would be around 13000 times and 40\% if two millions of cores were employed.

\begin{figure}
\centering
\includegraphics[width=3.2in]{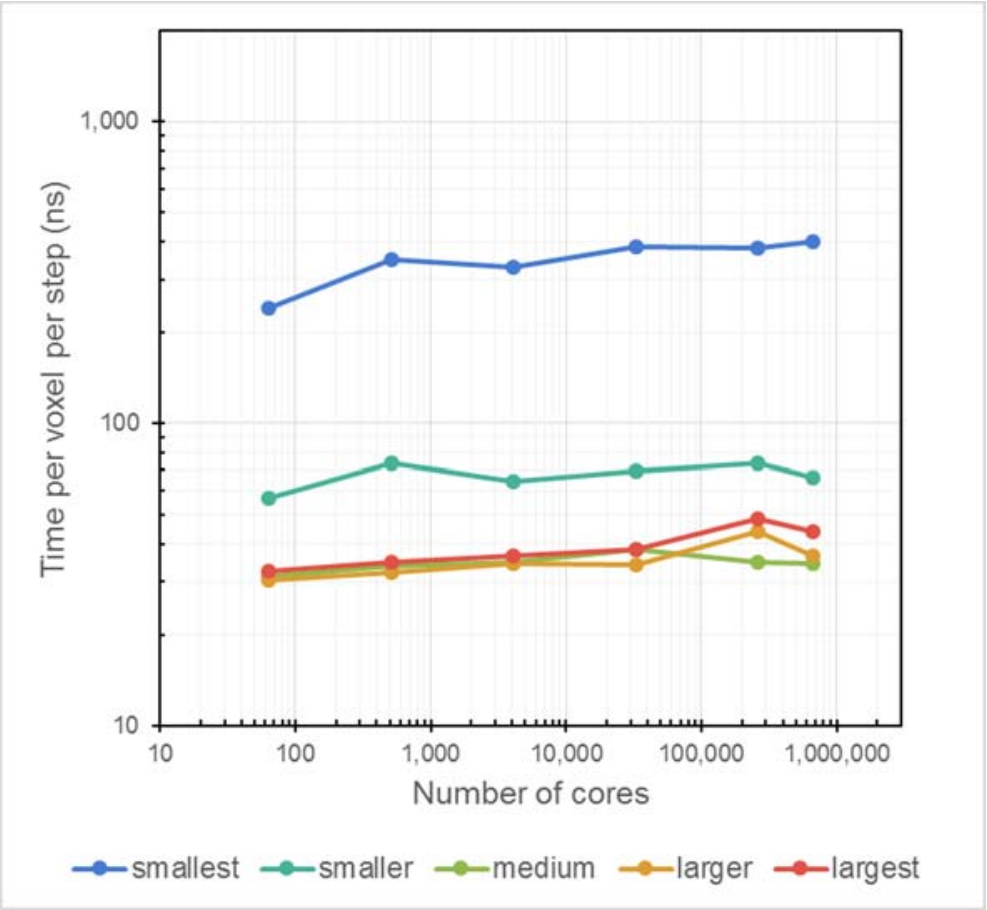}
\caption{Elapsed times according to lattice size per core. Smallest to largest lines represent $32^3$, $64^3$, $128^3$, $256^3$, and $512^3$ voxels per process, respectively.}
\label{fig:weak}
\end{figure}

\begin{figure}
\centering
\includegraphics[width=3.2in]{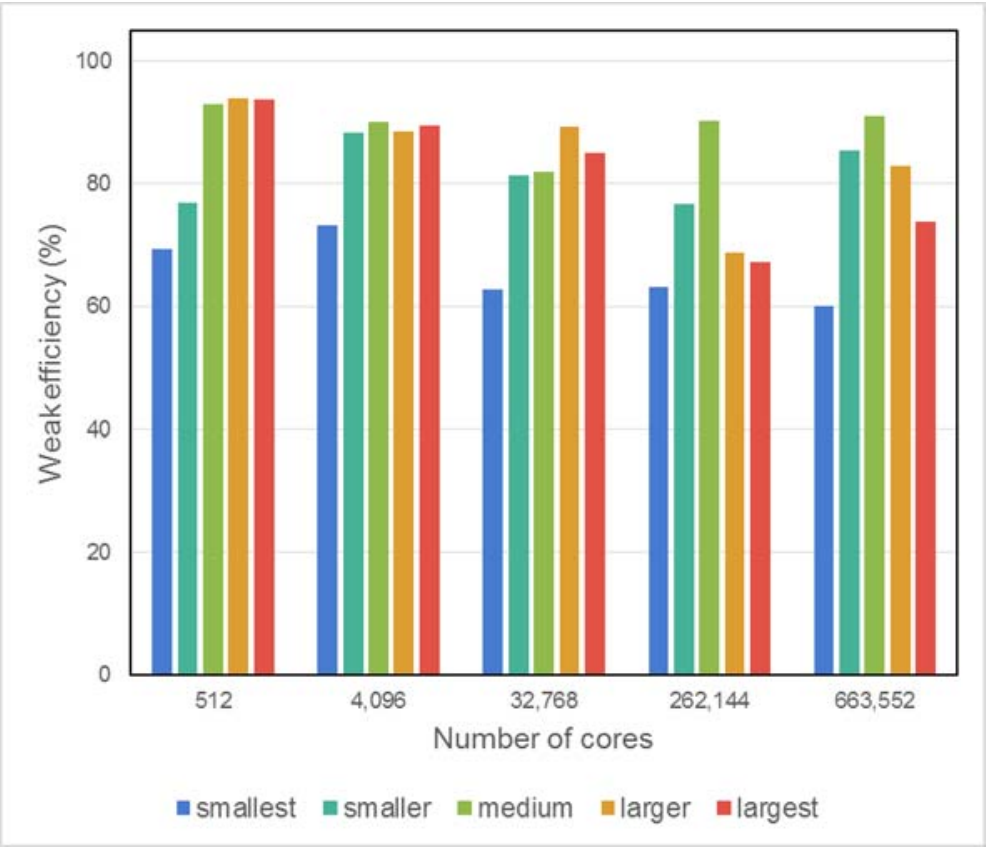}
\caption{Weak efficiency. Colors are the same as those in Fig. \ref{fig:weak}.}
\label{fig:weakefficiency}
\end{figure}

In terms of weak (i.e., extensive) scaling, Fig. \ref{fig:weak} shows the elapsed time per voxel per step. Labels, smallest, smaller, medium, larger or largest, in the figure indicate that the lattice on each node is $16^3$, $32^3$, $64^3$, $128^3$ or $256^3$. For any lattice size, those times should be identical (i.e. independent of the number of cores) if weak scaling is completely accomplished. In spite of variation in absolute value, all data sequences in the figure show good scaling properties. 
As for smallest and smaller lattices, scrutinizing the element of elapsed time revealed that constant time originated from latency overwhelmed calculations. This explains the cause of larger time in absolute value compared to the other lattices. Parallel efficiencies for weak scaling are summarized in Fig. \ref{fig:weakefficiency}. In the figure, elapsed time of 64 cores was used as the denominator for each lattice. Although efficiencies tend to deteriorate for high number of cores, at least 60\% of efficiency was accomplished up to more than a half million cores.

\begin{figure}
\centering
\includegraphics[width=3.2in]{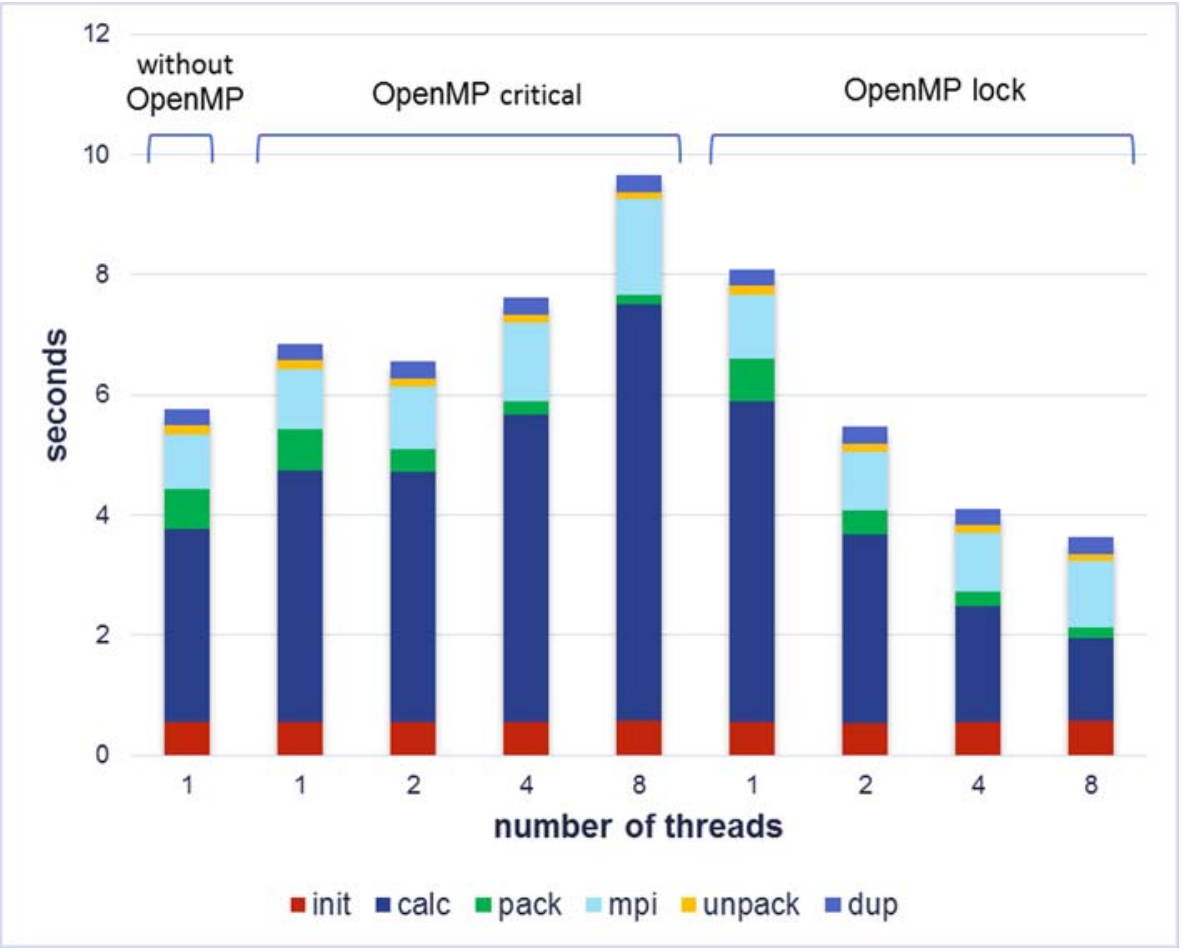}
\caption{Thread parallel performance, Elapsed times with and without OpenMP parallelization.}
\label{fig:thread}
\end{figure}

Elapsed times of $160^3$ voxels with 96 cores are shown in Fig. \ref{fig:thread}. Molecular occupancy is set as 0.3 and 1000 steps are calculated. Here the focus is on the effect of thread parallelization. Here the focus should be on the effect of thread parallelization so that MPI times are not important. With OpenMP critical function, pack time decreases according to the number of cores while calculation time increases. After all, total elapsed time increases. On the other hand, object locking technique shows a decrease in both pack and calculation times. However, there is an obvious saturation in performance as seen in the figure. Apart from init and MPI times, the remaining time with eight threads is just four times faster than that with a single thread. Considering actual contention rarely occurs for this occupancy of molecules, we suspect the reason of underperformance is due to the overheads inherent in OpenMP.

\subsection{MAPK model}
As a preliminary example of real-world application, we performed parallel calculation of the MAPK model, a typical network motif in cell signaling pathways. Because complete cell signaling pathways are very huge networks \cite{mapk:1}, we made the pathway as simple as possible without any loss of essential parts. In the MAPK model, MAPK is doubly phosphorylated by MAPK kinase (MAPKK) and doubly dephosphorylated by phosphatase (Pase). This model consists of nine chemical species and twelve reactions as formulated in the following. 
\begin{eqnarray*} 
\mathrm{MAPK} + \mathrm{MAPKK} &\overset{k_1}{\underset{k_2}{\rightleftarrows}}& \mathrm{MAPK/MAPKK}, \cr
\mathrm{pMAPK} + \mathrm{MAPKK} &\overset{k_4}{\underset{k_5}{\rightleftarrows}}& \mathrm{pMAPK/MAPKK}, \cr
\mathrm{ppMAPK} + \mathrm{Pase} &\overset{k_1}{\underset{k_2}{\rightleftarrows}}& \mathrm{ppMAPK/Pase}, \cr
\mathrm{pMAPK} + \mathrm{Pase} &\overset{k_4}{\underset{k_5}{\rightleftarrows}}& \mathrm{pMAPK/Pase}, \cr
\mathrm{MAPK/MAPKK} &\overset{k_3}{\rightarrow}& \mathrm{pMAPK} + \mathrm{MAPKK}, \cr
\mathrm{pMAPK/MAPKK} &\overset{k_6}{\rightarrow}& \mathrm{ppMAPK} + \mathrm{MAPKK}, \cr
\mathrm{ppMAPK/Pase} &\overset{k_3}{\rightarrow}& \mathrm{pMAPK} + \mathrm{Pase}, \cr
\mathrm{pMAPK/Pase} &\overset{k_6}{\rightarrow}& \mathrm{MAPK} + \mathrm{Pase}. 
\end{eqnarray*}
The initial conditions and the kinetic parameters were $k_1 = 0.027\,nM^{-1}s^{-1}$, $k_2 = 1.3\,s^{-1}$, $k_3 = 1.5\,s^{-1}$, $k_4 = 0.026\,nM^{-1}s^{-1}$, $k_5 = 1.73\,s^{-1}$, $k_6 = 15.0\,s^{-1}$, $D = 1\,\mu m^2/s$ and $r_v = 10 \,nm$. The initial concentrations of MAPK, MAPKK and Pase were$ 200\,nM$, $50\,nM$ and $50\,nM$, respectively. The lattice size and its physical volume were $480^3$ and $0.443\,pL$.
In the chemical formulas above, the prefixes p and pp stand for singly and doubly phosphorylated MAPK, respectively. 

\begin{figure}
\centering
\includegraphics[width=3.4in]{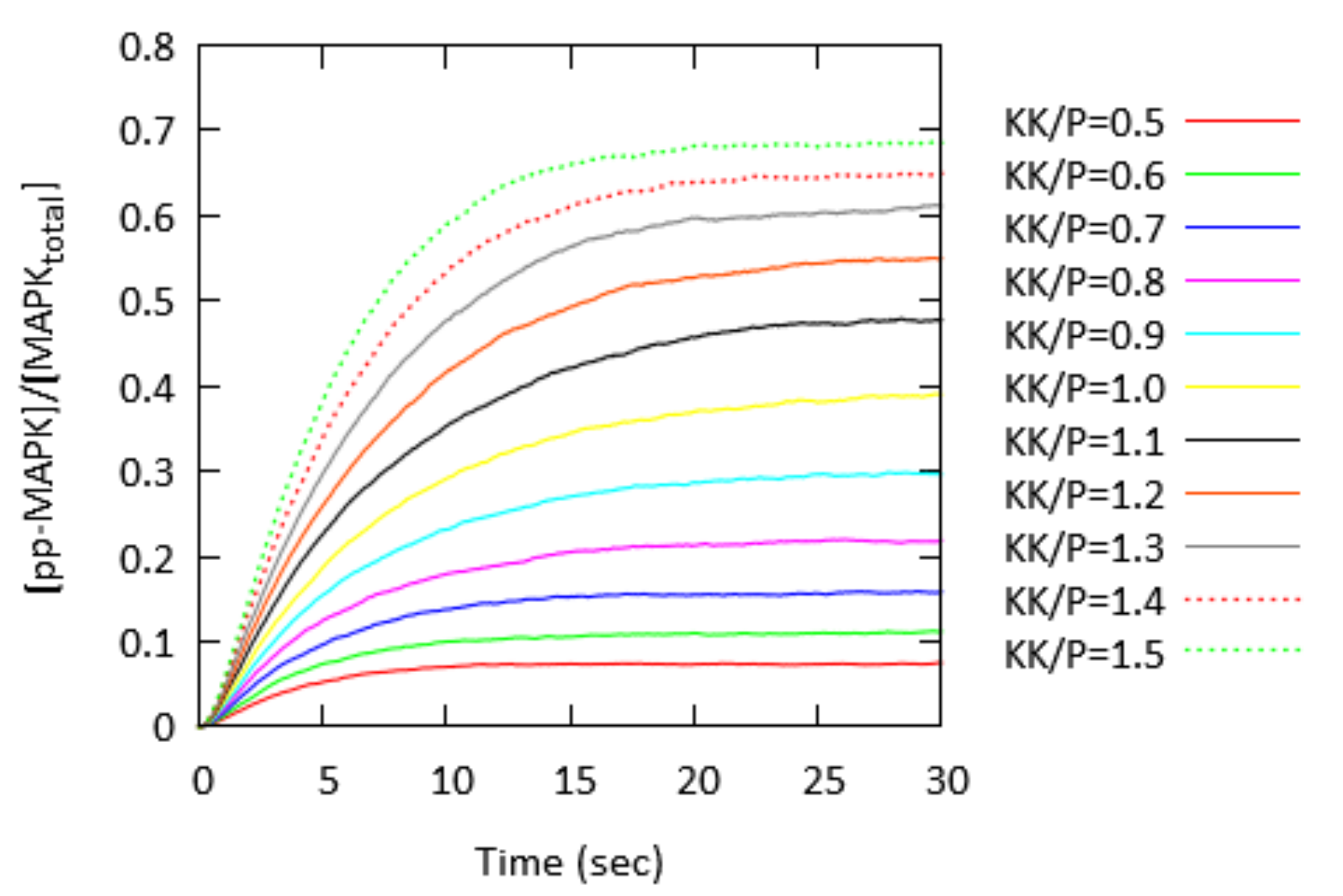}
\caption{Simulated results of a MAPK model. Fractions of ppMAPK are shown. KK and P denote MAPK kinase and phosphatase, respectively. Phosphatase concentration is fixed at 50 nM. Colored lines represent simulated results for different initial concentrations of MAPK kinase (25-75 nM).}
\label{fig:mapk}
\end{figure}

It is a rather simple model, and the parameters are based on experimental data from \cite{mapk:2}. In the calculation, the molecules of all species were assumed to be initially distributed homogeneously for simplicity. All calculations were carried out with 4,096 cores (2,048 processes, two threads per process) and required approximately 16 hours (57,600 seconds). Figure \ref{fig:mapk} shows ppMAPK fractions relative to total MAPK concentrations during a thirty-second period for various initial fractions of phosphatase. Although there is still room for quantitative scrutiny, the dependence on the parameters and the behavior of the variation agree with experimental results at least qualitatively. Calculations with larger pathways, longer durations, and membrane structures are under way.

\section{Conclusions}
We proposed a parallel stochastic method to calculate cell signaling in a hexagonal close-packed lattice. To realize large-scale parallel calculations, several improvements, including a twisted Cartesian coordinate system, an event scheduler, random sub-volumes and a voxel-locking technique, were introduced. We also validated the physical correctness of the program. The calculated diffusion rates without crowding coincided with theory within negligibly small error. Under the crowding condition, the diffusion rates decreased, as expected by theory. The first-order reaction agreed well with theory, and the second-order reaction agreed with our well-validated serial program. Parallel performance was sufficiently high for large-scale calculations. From the viewpoint of strong scaling, our program achieved 7686 times speedup in the case of 663552 cores relative to 64 cores with a $2048^3$ lattice. Its efficiency was equal to 74.1\%. In terms of weak scaling, efficiencies at least 60\% were obtained. Voxel-locking technique was shown to work well to parallelize randomly accessing loops. In addition, a preliminary calculation of the MAPK model revealed that the program is promising for application to real problems. Last but not least, twisted Cartesian coordinate also enables calculations of three dimensional lattice gas automata and D3Q13 model in lattice Boltzmann methods besides conventional models such as D3Q15, D3Q19 or D3Q27.


\section{Acknowledgments}
This research is part of the HPCI Strategic Program for Innovative Research in Supercomputational Life Science (SPIRE Field 1), which is funded by the Ministry of Education, Culture, Sports, Science and Technology (MEXT), Japan. Part of the results were obtained by using the K computer at RIKEN Advanced Institute for Computational Science (Proposal number hp120309). We would like to thank Yukihiro Eguchi for continuous encouragement; and Peter Karagiannis and Kylius Wilkins of QBiC for reading the manuscript and making valuable comments. Finally, first auther wish to thank Hisashi Nakamura of RIST for giving him the opportunity to start this research.


\bibliographystyle{abbrv}
%
%

\newpage
\onecolumn

\begin{appendix}

\bigskip

\section{Configuration of HCP lattice.}

\begin{figure}[htbp]
\centering
\includegraphics[width=\linewidth]{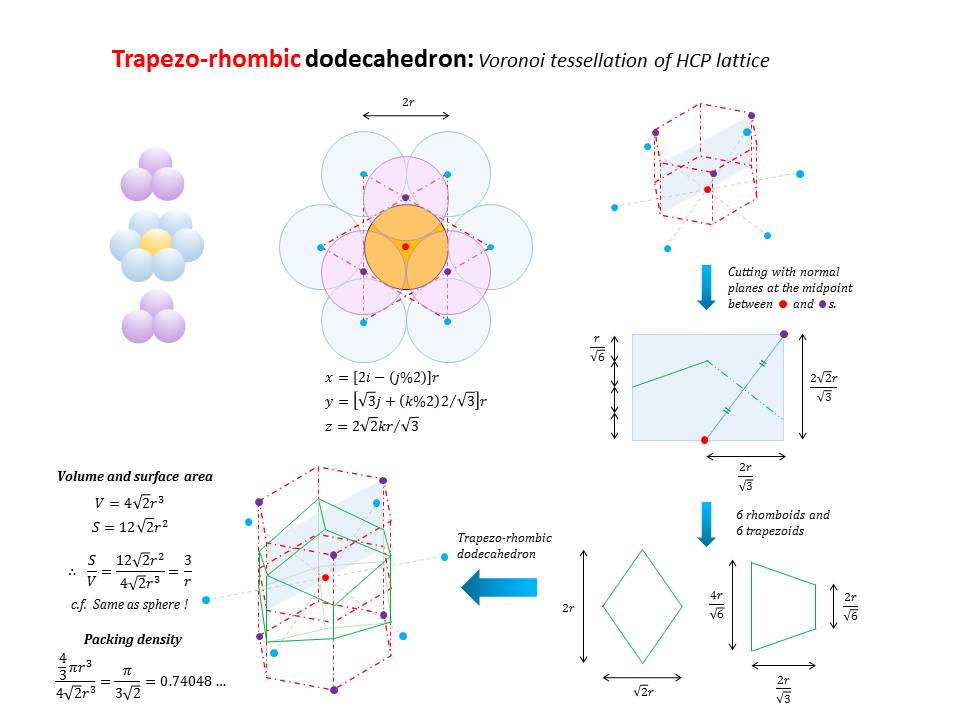}
\end{figure}

\begin{figure}[htbp]
\centering
\includegraphics[width=1.0\linewidth]{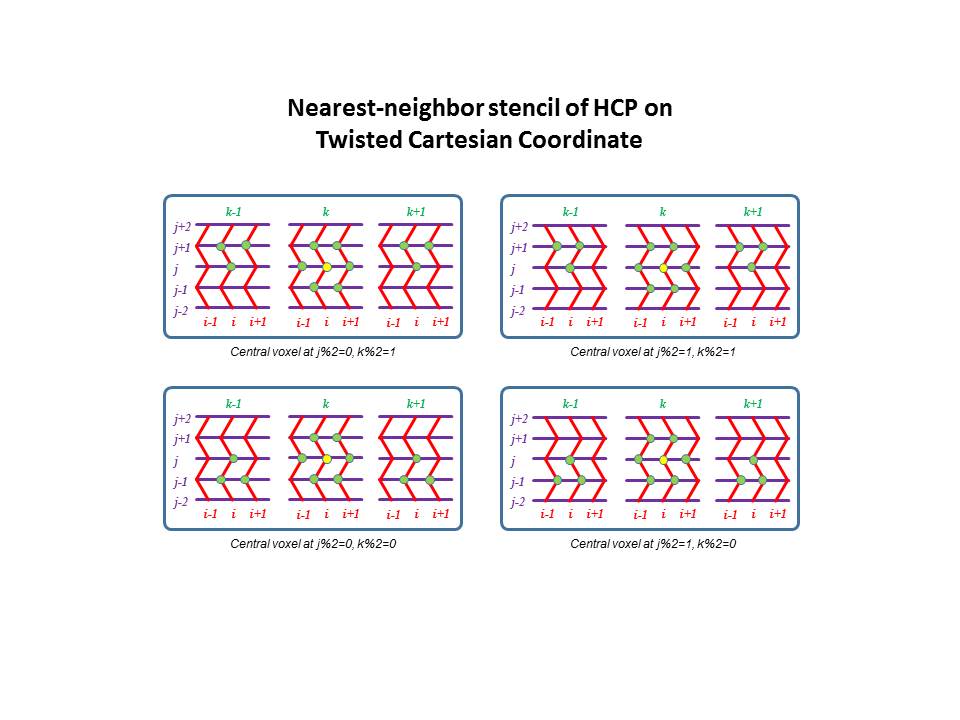}
\end{figure}

\begin{figure}[htbp]
\centering
\includegraphics[width=0.9\linewidth,trim=10mm 30mm 10mm 30mm,clip]{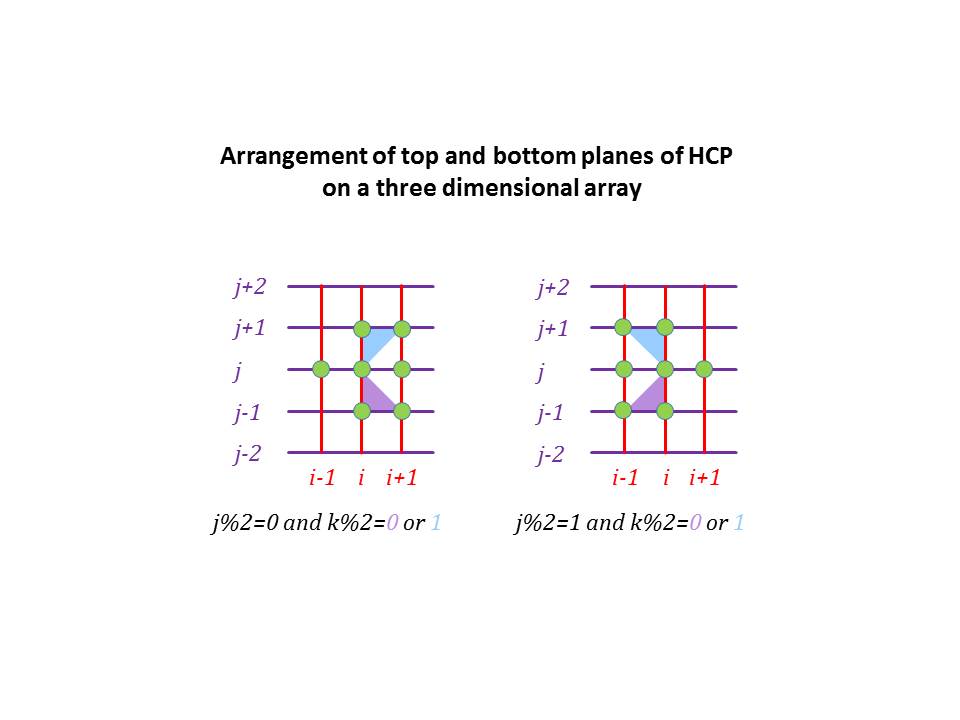}
\end{figure}

\vfill
\newpage

\section{Configuration of FCC lattice.}
\begin{figure}[htbp]
\centering
\includegraphics[width=\linewidth]{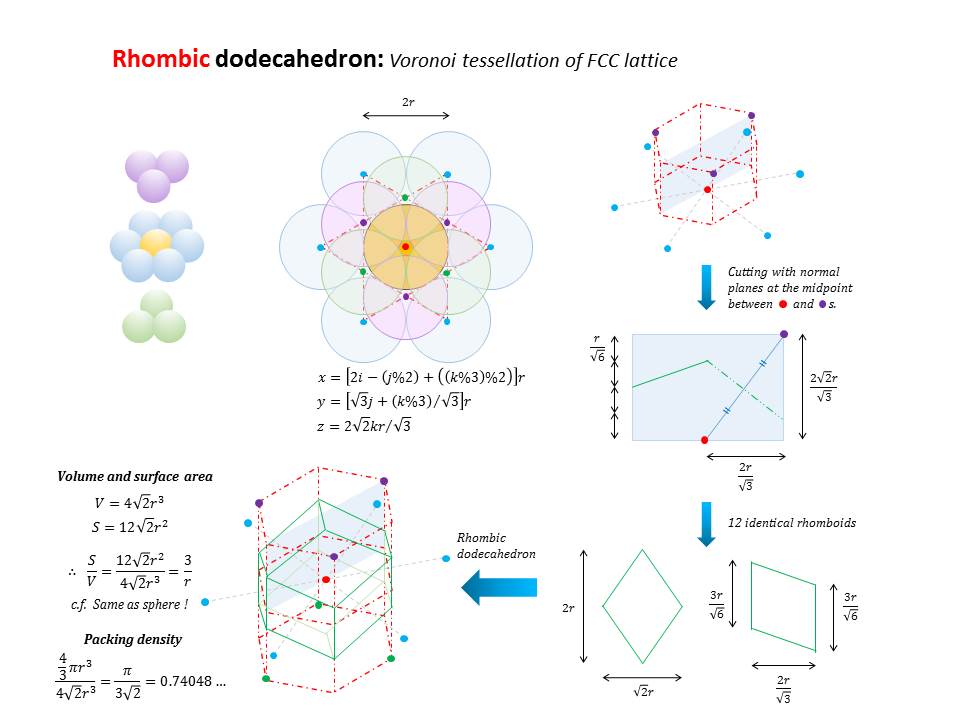}
\end{figure}

\begin{figure}[htbp]
\centering
\includegraphics[width=1.0\linewidth]{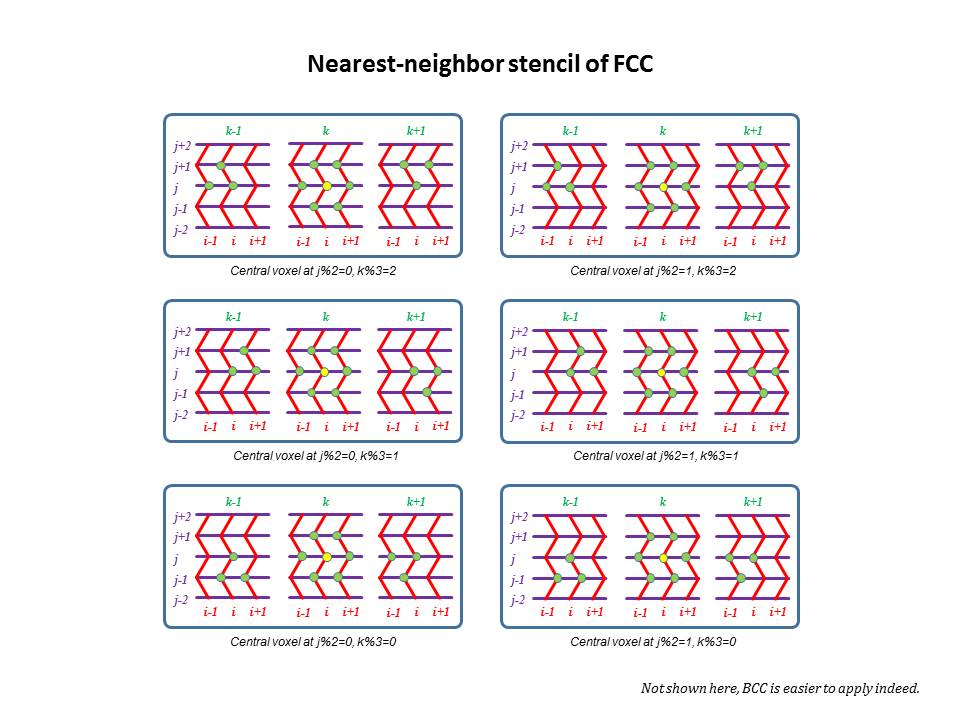}
\end{figure}

\begin{figure}[htbp]
\centering
\includegraphics[width=0.9\linewidth,trim=0mm 30mm 0mm 30mm,clip]{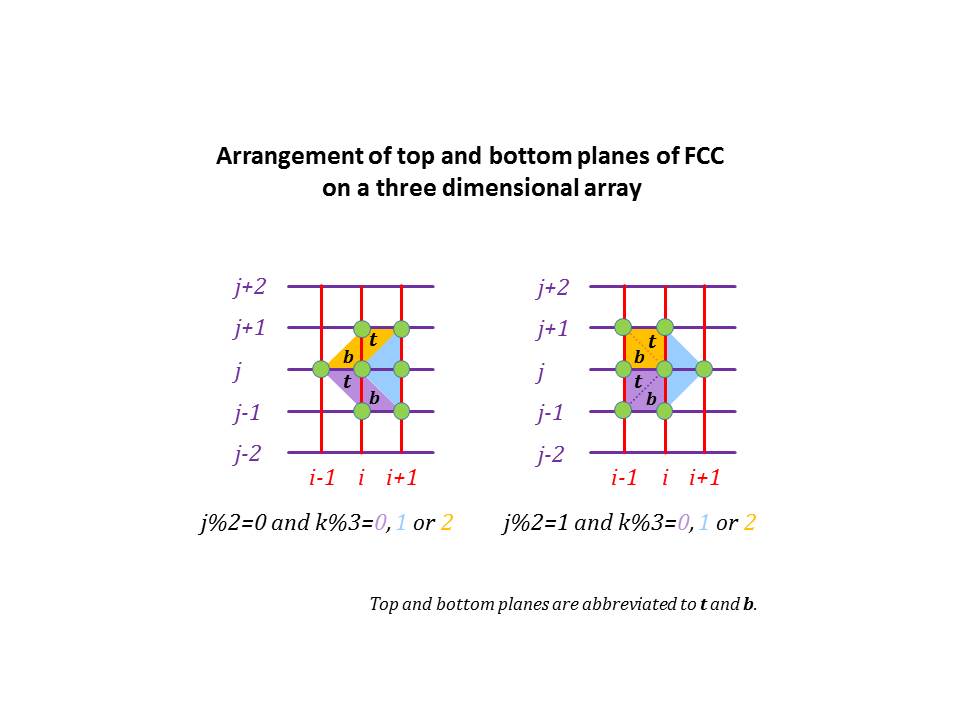}
\end{figure}

\newpage

\section{A simple derivation of the Einstein-Smoluchowski equation.}
\bigskip
Mathematically, an infinitesimal limit of a random walk can be formulated as a Wiener process. The traveling distance $d{\bf X}_t$ in three-dimensional space in time interval $dt$ can be written in general by a stochastic differential equation (SDE) as 
\begin{equation} d{\bf X}_t= {\boldsymbol \mu} dt+\sigma d{\bf W}_t, \label{eq:sde1} \end{equation}
where ${\boldsymbol \mu}$, $\sigma$, and ${\bf W}_t$ denote drift, variance, and the three-dimensional Wiener process, respectively. Characters in boldface indicate three-dimensional vectors. The Wiener process is a mathematically defined stochastic variable that starts from the origin and obeys a normal distribution. In the case of the diffusion equation, drift and variance parameters are considered constants. Without any loss of generality, we can assume that ${\bf X}_0 = {\bf 0}$. Consequently, eq. (\ref{eq:sde1}) can be integrated in the sense of the Ito integral as 
\begin{equation} {\bf X}_t = {\boldsymbol \mu} t+\sigma {\bf W}_t. \label{eq:sde2} \end{equation}
Note that the average and the covariance of the three-dimensional Wiener process are defined as: 
\begin{equation} E[{\bf W}_t ] = {\bf 0}, \quad E[{\bf W}_t^2 ] = 3t, \label{eq:average1} \end{equation}
where $E$ stands for expectation over the ensembles. Applying eq. (\ref{eq:average1}) to eq. (\ref{eq:sde2}), the average and the covariance of ${\bf X}_t$ can be readily found, 
\begin{equation}E[{\bf X}_t ] = {\boldsymbol \mu} t, \quad E[{\bf X}_t^2 ] = \mu^2 t^2 + 3\sigma^2 t. \label{eq:expect} \end{equation}
Since $t$ is a deterministic variable, it does not affect the expectation. 

On the other hand, macroscopically, 
the Kolmogorov forward equation, sometimes known as the Fokker-Planck equation, is also formally derived from the given SDE as 
\begin{equation} {\partial_t p} = -{\nabla }\cdot\left({\boldsymbol \mu} p \right) + {1 \over 2} 
\triangle\left (\sigma^2 p \right), \end{equation}
where $p$ denotes a probabilistic distribution that evolves from the initial distribution of molecules \cite{shreve:1}. Comparing this equation with the diffusion equation considered in this paper, 
\begin{equation} {\partial_t p} = D {\triangle p}, \end{equation}
we immediately find the following relations, 
\begin{equation} {\boldsymbol \mu} = {\bf 0}, \quad D = \sigma^2 / 2. \label{eq:coef} \end{equation}
Substituting eq. (\ref{eq:coef}) into eq. (\ref{eq:expect}) gives 
\begin{equation}E[{\bf X}_t ] = {\bf 0}, \quad E[{\bf X}_t^2 ] = 6Dt. \label{eq:xt2} \end{equation}
Because $E[{\bf X}_t^2 ]$ has dimensions in squared length, we denote it by $\lambda$. Finally, we can rewrite the latter of eq. (\ref{eq:xt2}) as 
\begin{equation} \lambda^2 = 6Dt. \label{eq:msd} \end{equation}
This equation is called the Einstein-Smoluchowski equation.

\newpage

\section{An example of simulation pipeline.}

\begin{figure}[htbp]
\centering
\includegraphics[width=\linewidth]{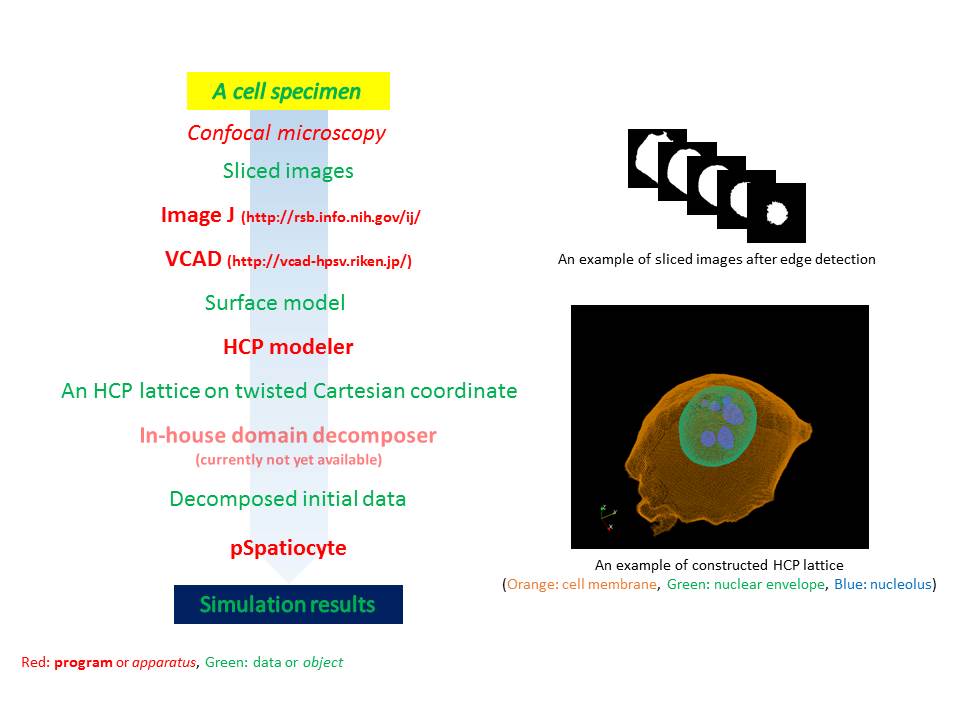}
\end{figure}

\end{appendix}

\balancecolumns
\end{document}